\documentclass[a4paper,12pt]{article}
\pdfoutput=1
\usepackage{amssymb}
\usepackage{amsmath}
\usepackage{amsfonts}
\usepackage{mathtools}
\usepackage{slashed}
\usepackage{color}
\usepackage{indentfirst}
\usepackage{graphicx}
\usepackage{csquotes} 
\usepackage{caption}
\usepackage{cite}
\usepackage{here}
\usepackage{comment}
\usepackage{listings}
\usepackage{booktabs}
\usepackage[italicdiff]{physics}
\numberwithin{equation}{section}

\makeatletter
\newcommand*\rel@kern[1]{\kern#1\dimexpr\macc@kerna}
\newcommand*\widebar[1]{%
  \begingroup
  \def\mathaccent##1##2{%
    \rel@kern{0.8}%
    \overline{\rel@kern{-0.8}\macc@nucleus\rel@kern{0.2}}%
    \rel@kern{-0.2}%
  }%
  \macc@depth\@ne
  \let\math@bgroup\@empty \let\math@egroup\macc@set@skewchar
  \mathsurround\z@ \frozen@everymath{\mathgroup\macc@group\relax}%
  \macc@set@skewchar\relax
  \let\mathaccentV\macc@nested@a
  \macc@nested@a\relax111{#1}%
  \endgroup
}
\makeatother

\renewcommand{\thefootnote}{\fnsymbol{footnote}}

\begin{document}

\title{
\begin{flushright}
\begin{minipage}{0.2\linewidth}
\normalsize
EPHOU-21-010\\
KEK-TH-2333 \\*[50pt]
\end{minipage}
\end{flushright}
{\Large \bf 
Symplectic modular symmetry in heterotic string vacua:\\
flavor, CP, and $R$-symmetries
\\*[20pt]}}

\author{Keiya Ishiguro$^{a}$\footnote{
E-mail address: ishigu@post.kek.jp
},\,
Tatsuo Kobayashi$^{b}$\footnote{
E-mail address: kobayashi@particle.sci.hokudai.ac.jp
}
\ and\
Hajime~Otsuka$^{c}$\footnote{
E-mail address: hotsuka@post.kek.jp
}\\*[20pt]
$^a${\it \normalsize 
Graduate University for Advanced Studies (Sokendai),}\\
{\it \normalsize 1-1 Oho, Tsukuba, Ibaraki 305-0801, Japan.} \\
$^b${\it \normalsize 
Department of Physics, Hokkaido University, Sapporo 060-0810, Japan} \\
$^c${\it \normalsize 
KEK Theory Center, Institute of Particle and Nuclear Studies, KEK,}\\
{\it \normalsize 1-1 Oho, Tsukuba, Ibaraki 305-0801, Japan}}
\maketitle

\date{
\centerline{\small \bf Abstract}
\begin{minipage}{0.9\linewidth}
\medskip 
\medskip 
\small
We examine a common origin of four-dimensional flavor, CP, and $U(1)_R$ symmetries 
in the context of heterotic string theory with standard embedding. 
We find that flavor and $U(1)_R$ symmetries are unified into 
the $Sp(2h+2, \mathbb{C})$ modular symmetries of Calabi-Yau threefolds with $h$ being the number of moduli fields. 
Together with the $\mathbb{Z}_2^{\rm CP}$ CP symmetry, they are enhanced to $GSp(2h+2, \mathbb{C})\simeq Sp(2h+2, \mathbb{C})\rtimes \mathbb{Z}_2^{\rm CP}$ generalized symplectic modular symmetry. 
We exemplify the $S_3, S_4, T^\prime, S_9$ non-Abelian flavor symmetries on explicit toroidal orbifolds with and 
without resolutions and $\mathbb{Z}_2,S_4$ flavor symmetries on three-parameter examples of Calabi-Yau threefolds. 
Thus, non-trivial flavor symmetries appear in not only the exact orbifold limit but also a certain class of Calabi-Yau threefolds.
These flavor symmetries are further enlarged to non-Abelian discrete groups by the CP symmetry.

\end{minipage}
}

\renewcommand{\thefootnote}{\arabic{footnote}}
\setcounter{footnote}{0}
\thispagestyle{empty}
\clearpage
\addtocounter{page}{-1}

\tableofcontents

\section{Introduction}
\label{sec:1}

The origin of flavor and CP symmetries is one of the unsolved and fundamental issues in the Standard Model. 
Understanding the underlying structure would give us a clue to find out the ultra-violet 
completion of the Standard Model. 
It was proposed in the pioneering works \cite{Feruglio:2017spp} that these symmetries are closely connected with the geometrical symmetry of extra-dimensional spaces. 
In particular, $SL(2,\mathbb{Z})$ modular symmetry is of particular interest in the bottom-up 
approach because the non-Abelian discrete flavor symmetries, which are used to explain flavor structure of quarks and leptons 
such as $A_4$, $S_4$, $A_5$ in the past decades, are naturally 
embedded into the $SL(2,\mathbb{Z})$, called the modular flavor symmetry \cite{deAdelhartToorop:2011re}. 
The inclusion of the CP symmetry enlarges  $SL(2,\mathbb{Z})$ to the generalized modular group $GL(2,\mathbb{Z})$ \cite{Nilles:2018wex,Novichkov:2019sqv}.

From the viewpoint of top-down approach such as the string theory, the $SL(2,\mathbb{Z})$ symmetry appears in the low-energy effective action from the higher-dimensional theories on toroidal backgrounds \cite{Ferrara:1989bc,Ferrara:1989qb} \footnote{
Also, the modular symmetry, in particular, its anomaly was studied in Refs.~\cite{Derendinger:1991hq,Ibanez:1992hc,Kobayashi:2016ovu}.}. 
Furthermore, the flavor symmetry can be a quotient of $SL(2,\mathbb{Z})$, 
and CP symmetry can be regarded as an outer automorphism of $SL(2,\mathbb{Z})$.
They can be treated in a unified manner \cite{Baur:2019kwi}. 
The stabilization mechanism of moduli values providing observed masses and mixing angles of quarks/leptons is proposed in the context of string theory \cite{Kobayashi:2020hoc,Kobayashi:2020uaj,Ishiguro:2020nuf}.
Hence, both flavor and CP symmetries have a common origin in the geometrical symmetries of toroidal backgrounds, 
and this approach would be useful to reveal the nature of extra-dimensional spaces relevant to particle physics.

In this paper, we examine the existence of modular flavor symmetry on Calabi-Yau (CY) threefolds which are promising backgrounds of the string theory. 
The complex structure and K\"ahler moduli spaces of CY threefolds are described by the symplectic structure with 
$Sp(2h+2, \mathbb{Z})$ symplectic groups.\footnote{We refer the details about the symplectic modular symmetry to Refs. \cite{Strominger:1990pd,Candelas:1990pi}.} 
Here, $h$ is determined by the number of the complex structure moduli $h=h^{2,1}$ or the K\"ahler moduli $h=h^{1,1}$. 
Such a symplectic extension was recently discussed in Refs. \cite{Ding:2020zxw,Ishiguro:2020nuf,Baur:2020yjl,Nilles:2021glx} in both bottom-up and top-down approaches, and these flavor symmetries are called the symplectic modular flavor symmetries. 
Since moduli fields and matter fields are in one-to-one correspondence with each other in the context of heterotic string theory with standard embedding, the flavor symmetries of matter fields naturally lie in symplectic modular flavor symmetries. 
In particular, it turned out that four-dimensional (4D) CP symmetry was regarded as the outer automorphism of $Sp(2h+2,\mathbb{Z})$ on CY backgrounds \cite{Ishiguro:2020nuf}. 
These observations motivate us to examine the symplectic modular symmetry involving the flavor and CP symmetries in the context of CY symplectic moduli spaces. 

The purpose of this paper is to formulate the symplectic modular flavor symmetries together with CP symmetry in the context of heterotic string theory with standard embedding and exemplify these symmetries on several toroidal and CY backgrounds. 
As discussed in detail in Sec. \ref{sec:2}, the flavor symmetries of fundamental and anti-fundamental representations of $E_6$ gauge group are governed by the symplectic modular symmetries in a general class of CY threefolds with large volume/complex structure, although the flavor structure depends on the intersection number of CY threefolds. 
Together with the results of Ref. \cite{Ishiguro:2020nuf}, we find that symplectic modular flavor symmetries are further enlarged to $GSp(2h+2, \mathbb{Z})\simeq Sp(2h+2, \mathbb{Z})\rtimes \mathbb{Z}_2^{\rm CP}$, called ``generalized symplectic modular symmetry''. 
Furthermore, the rotation of holomorphic three-form of CY threefolds induces that of Killing spinors of  4D supersymmetry, namely the $U(1)_R$ symmetry \cite{Witten:1985xc}. 
As a result, the flavor and the $U(1)_R$ symmetries are described by $Sp(2h+2, \mathbb{C})$ symplectic modular symmetry, which results in the unification of all flavor, CP, and $U(1)_R$ symmetries in the context of $GSp(2h+2, \mathbb{C})$ symmetry. 
We explicitly demonstrate generalized symplectic modular symmetries on toroidal orbifolds with/without resolutions and three-parameter examples of CY threefolds. 
It turns out that there exist $G_{\rm flavor}=\mathbb{Z}_2, S_3, S_4, T^\prime, S_9$ non-Abelian flavor symmetries, and they are further enhanced into $G_{\rm flavor}\rtimes \mathbb{Z}_2^{\rm CP}$ together with the CP symmetry. 
So far, several techniques have been developed to understand $G_{\rm flavor}$ on toroidal orbifolds, but our approach is useful to find out the flavor structure of smooth CY threefolds, including the known toroidal orbifold regime. 
Indeed, our results indicate that a certain class of CY threefolds has the $S_4$ flavor symmetry on matter fields associated with the moduli fields. 

This paper is organized as follows. 
In Sec. \ref{sec:2}, we briefly review the symplectic structure of moduli fields in CY compactifications and propose the generalized symplectic modular symmetries of matter fields in the context of heterotic string theory with standard embedding, where fundamental and anti-fundamental representations of $E_6$ group are directly related to K\"ahler moduli and complex structure moduli, respectively. 
The symplectic modular flavor symmetries and their enlarged symmetries by the CP transformation are explicitly demonstrated in Sec. \ref{sec:3}. 
Sec. \ref{sec:con} is devoted to conclusions and discussions. 
In Appendix \ref{app}, we give the modular form of weight 1 for $T'$.

\section{Generalized symplectic modular symmetry}
\label{sec:2}

In Sec. \ref{sec:symplectic}, we review the symplectic structure of moduli fields in CY compactifications with an emphasis on 
the large volume/complex structure regime. 
In Sec. \ref{sec:modularflavor}, we give a prescription of the symplectic modular flavor symmetry of matter fields associated with the complex structure moduli and K\"ahler moduli in the context of heterotic string theory with standard embedding. 
Furthermore, the symplectic transformations of matter fields naturally incorporate the existence of $U(1)_R$ 
symmetry appearing in the 4D effective action as shown in Sec. \ref{sec:U(1)R}. 
The CP transformations exchanging the fundamental and anti-fundamental representations of $E_6$ are regarded as the anti-holomorphic transformations of moduli fields. 
In Sec. \ref{sec:unification}, we find that these anti-holomorphic transformations enlarge the symplectic modular symmetry into the generalized symplectic modular symmetry. 
As a result, the symplectic modular, flavor, CP, and $U(1)_R$ symmetries are treated in a unified manner.

\subsection{Symplectic structure of moduli fields in Calabi-Yau compactifications}
\label{sec:symplectic}

It was well known that the CY compactifications lead to the symplectic structure 
in the 4D effective action. (For more details about the symplectic structure of CY moduli spaces, see, e.g., Refs. \cite{Strominger:1990pd,Candelas:1990pi}.) 
The moduli fields associated with the CY metric are categorized into two types of moduli fields, i.e., the 
complex structure moduli and the K\"ahler moduli. 
Their K\"ahler potentials in the units of the reduced Planck mass are given by
\begin{align}
    K_{\rm cs} &= - \ln \biggl[-i \int_{{\cal M}} \Omega \wedge \widebar{\Omega}\biggl],
    \nonumber\\
    K_{\rm ks} &= - \ln \biggl[\int_{\cal M} J_c \wedge J_c \wedge J_c\biggl],
    \label{eq:KcsKks}
\end{align}
where $\Omega$ and $J_c$ denote the holomorphic three-form and (complexified) K\"ahler form of CY threefolds ${\cal M}$, respectively. 
Here, $J_c=B+iJ$ consists of real Kalb-Ramond two-form $B$ and K\"ahler form $J$. 
Their moduli spaces possess the symplectic structure. For concreteness, we will focus on the complex structure moduli with large complex structure, paying attention to the symplectic transformations of the moduli fields and the prepotential in subsequent discussions. 
The purpose of reviewing the symplectic structure is to understand the flavor structure of matter fields in the context of heterotic string theory with standard embedding, as will be discussed in the next section. 

The holomorphic three-form $\Omega$ is 
expanded on the symplectic basis $(\alpha_I, \beta^I)$ of $H^3({\cal M},\mathbb{Z})$,\footnote{Note that the $Sp(2h^{2,1}+2,\mathbb{C})$ appears in the classical moduli space, but loop and quantum corrections break it to $Sp(2h^{2,1}+2,\mathbb{Z})$, namely duality transformations of the string theory.} 
\begin{align}
    \Omega =\sum_{I=0}^{h^{2,1}} \left(X^I \alpha_I - {\cal F}_I \beta^I \right),
\end{align}
where $X^I$ are projective coordinates and the structure of ${\cal F}_I=\partial_I {\cal F}$ is determined by the prepotential ${\cal F}$. 
The orthogonal relation of the symplectic basis $(\alpha_I, \beta^I)$ 
\begin{align}
    \int_{\cal M} \alpha_I \wedge \beta^J = \delta^J_{\,\,\,\,I},
    \quad
    \int_{\cal M} \beta^J \wedge \alpha_I = -\delta^J_{\,\,\,\,I},
\end{align}
is invariant under the symplectic transformation:
\begin{align}
    \begin{pmatrix}
       \alpha_I\\
       \beta^I\\
    \end{pmatrix}
\rightarrow
    \begin{pmatrix}
       a & b\\
       c & d\\
    \end{pmatrix}
    \begin{pmatrix}
       \alpha_I\\
       \beta^I\\
    \end{pmatrix}
    ,
\end{align}
with
\begin{align}
        \begin{pmatrix}
       a & b\\
       c & d\\
    \end{pmatrix}
    \in Sp(2h^{2,1}+2,\mathbb{Z}).
\end{align}
Correspondingly, the vector $\{X^I, {\cal F}_I\}^T$ (the so-called period integrals of holomorphic three-form)
\begin{align}
    \Pi =
    \begin{pmatrix}
       \int_{A^I}\Omega \\
       \int_{B_I}\Omega\\
    \end{pmatrix}
    =
    \begin{pmatrix}
       X^I\\
       {\cal F}_I\\
    \end{pmatrix}
\end{align}
transforms as
\begin{align}
    \begin{pmatrix}
       X^I\\
       {\cal F}_I\\
    \end{pmatrix}
\rightarrow
    \begin{pmatrix}
       \widetilde{X}^I\\
       \widetilde{{\cal F}}_I\\
    \end{pmatrix}
=
        \begin{pmatrix}
       d & c\\
       b & a\\
    \end{pmatrix}
    \begin{pmatrix}
       X^I\\
       {\cal F}_I\\
    \end{pmatrix}
\label{eq:XFsymplecticTrf}
    ,
\end{align}
due to the fact that the holomorphic three-form is invariant under the symplectic transformation. 
Here, $\widetilde{{\cal F}}_I$ is required to be a derivative of new function $\widetilde{{\cal F}}(\widetilde{X})$, namely $\widetilde{{\cal F}}_I =\partial \widetilde{{\cal F}}(\widetilde{X})/ \partial \widetilde{X}^I$, to remain within the same class of 4D effective action. 
Note that the 4D ${\cal N}=2$ supersymmetric action is invariant under the symplectic transformation of the prepotential, 
$\widetilde{{\cal F}}(\widetilde{X}) = {\cal F}(X)$. 
Furthermore, when
\begin{align}
    \widetilde{{\cal F}}(\widetilde{X}) = {\cal F}(\widetilde{X}),
\end{align}
the Lagrangian of ${\cal N}=2$ supergravity with $n$ Abelian vector multiplets is invariant under the symplectic transformations \cite{deWit:1992wf}, although the prepotential itself is not invariant under them \cite{deWit:1984wbb}. 
This structure holds for the system of our interest, i.e., string compactifications on $(2,2)$ superconformal field theories \cite{Gross:1985fr,Gross:1985rr,Seiberg:1988pf,Cecotti:1988qn}. 

From the symplectic transformation (\ref{eq:XFsymplecticTrf}), it is found that $\{X^I\}$ takes a vector-valued modular form of $Sp(2h^{2,1}+2,\mathbb{Z})$, namely
\begin{align}
    X^I &\rightarrow \widetilde{X}^I= (c{\cal F}+d)^I_JX^J  =\frac{\partial \widetilde{X}^I}{\partial X^J}X^J,
\end{align} 
and the symmetric tensor ${\cal F}_{IJK}=\partial_I\partial_J\partial_K {\cal F}$ obeys 
\begin{align}
      {\cal F}_{IJK}&\rightarrow \widetilde{{\cal F}}_{IJK} = \frac{\partial X^L}{\partial \widetilde{X}^I}\frac{\partial X^M}{\partial \widetilde{X}^J}\frac{\partial X^N}{\partial \widetilde{X}^K}{\cal F}_{LMN},   
      \label{eq:FIJKTrf}
\end{align}
where it is required that $\partial \widetilde{X}^I/\partial X^J$ is a non-singular matrix. 
For instance, the invertible relation between $\widetilde{X}^I$ and $X^I$ is lost when the prepotential is a linear 
function of $X^I$ as pointed out in Ref. \cite{deWit:1984wbb}. 
This formula is useful to understand the symplectic transformation of Yukawa couplings among matter fields as analyzed in detail below. 

In the large complex structure regime\footnote{Throughout this paper, we analyze the classical CY moduli space.}, the prepotential is 
of the explicit form
\begin{align}
    {\cal F}(X) = \frac{1}{3!}\kappa_{ijk}\frac{X^iX^jX^k}{X^0} = (X^0)^2F(u),
\end{align}
where $\kappa_{ijk}$ denote the triple intersection numbers and the complex structure moduli $u^i$ are set by
\begin{align}
    u^i \equiv \frac{X^i}{X^0}.
\end{align}
Here and in what follows, we still call $F(u) = \frac{1}{3!}\kappa_{ijk}u^iu^ju^k$ 
the prepotential and take the flat coordinates 
\begin{align}
     X^0=1, \quad X^i = u^i.
     \label{eq:flatcoord}
\end{align}
Then, the symplectic transformation of the complex structure moduli $u^i$ is given by
\begin{align}
        u^i\equiv X^i &\rightarrow \widetilde{u}^i= \frac{\widetilde{X}^i}{\widetilde{X}^0}= \frac{\partial \widetilde{X}^i}{\partial X^J}\frac{X^J}{\widetilde{X}^0}= \frac{\partial \widetilde{X}^i}{\partial X^j}\frac{X^j}{\widetilde{X}^0}= \frac{\partial \widetilde{X}^i}{\partial X^j}\frac{u^j}{\widetilde{X}^0}
        = (c{\cal F}+d)^i_{j}\frac{u^j}{\widetilde{X}^0}.
        \label{eq:uTrans}
\end{align}
Let us discuss the symplectic transformation of Yukawa couplings ($F_{ijk}$) among complex structure moduli. 
They are determined by the triple intersection numbers $\kappa_{ijk}$,
\begin{align}
    F_{ijk} \equiv \pdv{u^i}\pdv{u^j}\pdv{u^k}F(u) =X^0 {\cal F}_{ijk}(X) = \kappa_{ijk},
\end{align}
which correspond to the holomorphic Yukawa couplings of matter fields in the context of heterotic string theory with standard embedding. 
The symplectic transformation of Yukawa couplings $F_{ijk}$ can be read off from Eq. (\ref{eq:FIJKTrf}), i.e.,
\begin{align}
        \widetilde{F}_{ijk} &=  \pdv{\widetilde{u}^i}\pdv{\widetilde{u}^j}\pdv{\widetilde{u}^k}\widetilde{F}(\widetilde{u}) = 
        \pdv{u^l}{\widetilde{u}^i}\pdv{u^m}{\widetilde{u}^j}\pdv{u^n}{\widetilde{u}^k}F_{lmn},
\label{eq:tildeFijk1}
\end{align}
or
\begin{align}
        \widetilde{F}_{ijk} &= \widetilde{X}^0\widetilde{{\cal F}}_{ijk} = \widetilde{X}^0\pdv{X^L}{\widetilde{X}^i}\pdv{X^M}{\widetilde{X}^j}\pdv{X^N}{\widetilde{X}^k}\,{\cal F}_{LMN}
        =   \widetilde{X}^0\pdv{X^l}{\widetilde{X}^i}\pdv{X^m}{\widetilde{X}^j}\pdv{X^n}{\widetilde{X}^k}\,F_{lmn}.
        \label{eq:tildeFijk2}
\end{align}
Hence, these two expressions lead to 
\begin{align}
    \pdv{u^l}{\widetilde{u}^i} = (\widetilde{X}^0)^{1/3}\pdv{X^l}{\widetilde{X}^i}.\quad
 \label{eq:tildeFijk3}
\end{align}

So far, we have focused on the symplectic structure of the complex structure moduli space, but the same structure appears in the K\"ahler moduli space as discussed in detail in Ref. \cite{Candelas:1990pi}. 
When we introduce a basis of $H^2({\cal M},\mathbb{Z})$ as $e_a$, the K\"ahler moduli $t^a$ are defined by
\begin{align}
    t^ae_a = B + i J =J_c,
\end{align}
whose prepotential in the large volume regime is given by
\begin{align}
   {\cal G}(Y)=\frac{1}{3!}\kappa_{abc} \frac{Y^aY^bY^c}{Y^0}= (Y^0)^2 G(t),
\label{eq:G}
\end{align}
with 
\begin{align}
    t^a = \frac{Y^a}{Y^0}.
\end{align}
Here, $\kappa_{abc}$ denote the triple intersection numbers among two cycles of CY threefolds, and $\{Y^A\}$ represent the projective coordinates on the K\"ahler moduli space. In particular, $Y^0$ is identified with the inverse of the axio-dilaton. 
Analogous to the complex structure moduli, the K\"ahler potential (\ref{eq:KcsKks}) is written by
\begin{align}
    K_{\rm cs} &= - \ln \biggl[-i \int_{{\cal M}} \Omega \wedge \widebar{\Omega}\biggl] = -\ln \biggl[-i (\widebar{X}^I{\cal F}_I -X^I\widebar{{\cal F}}_I) \biggl],
    \nonumber\\
    K_{\rm ks} &= - \ln \biggl[\int_{\cal M} J_c \wedge J_c \wedge J_c\biggl]= -\ln \biggl[-i (\widebar{Y}^A{\cal G}_A -Y^A\widebar{{\cal G}}_A) \biggl],
\end{align}
with $I=0,1,\cdots,h^{2,1}$, $A=0,1,\cdots,h^{1,1}$, and one can deal with the symplectic structure as summarized in Table \ref{tab:modulispaces}. 

\begingroup
\renewcommand{\arraystretch}{1.5}
\begin{table}[H]
   \begin{center}
   \scalebox{1.0}{
    \begin{tabular}{ccc} \toprule
       & Complex structure & K\"ahler structure \\ \midrule
      Projective coordinates & $X^I$ & $Y^A$\\ 
       & $(I=1,2,\cdots, h^{2,1}+1
      )$ & $(A=1,2,\cdots, h^{1,1}+1
      )$\\ 
      Moduli fields & $u^i$ & $t^a$\\ 
       & $(i=1,2,\cdots, h^{2,1}
      )$ & $(a=1,2,\cdots, h^{1,1}
      )$\\ 
      Prepotential & ${\cal F}(X) = (X^0)^2F(u)$ & ${\cal G}(Y) = (Y^0)^2G(t)$\\ 
      K\"ahler potential & $ -\ln \biggl[-i (\widebar{X}^I{\cal F}_I -X^I\widebar{{\cal F}}_I) \biggl]$ & $ -\ln \biggl[-i (\widebar{Y}^A{\cal G}_A -Y^A\widebar{{\cal G}}_A) \biggl]$\\
      Yukawa couplings & $\partial_i\partial_j\partial_k F = \kappa_{ijk}$ & $\partial_a\partial_b\partial_c G = \kappa_{abc}$\\ 
      Symplectic modular group & $Sp(2h^{2,1}+2,\mathbb{Z})$ & $Sp(2h^{1,1}+2,\mathbb{Z})$\\ 
      Symplectic transformations & $u^i \rightarrow (\widetilde{X}^0)^{-1} (c{\cal F}+d)^i_{j}u^j$ & $t^a \rightarrow (\widetilde{Y}^0)^{-1} (g{\cal G}+h)^a_{b}t^b$\\ 
       & $\begin{pmatrix}
            a & b\\
            c & d\\
          \end{pmatrix}
          \in Sp(2h^{2,1}+2,\mathbb{Z})
         $
       & $
          \begin{pmatrix}
            e & f\\
            g & h\\
          \end{pmatrix}
          \in Sp(2h^{1,1}+2,\mathbb{Z})
         $\\ \bottomrule
    \end{tabular}
    }
  \end{center}
  \caption{Moduli spaces of the complex structure and the K\"ahler structure.}
  \label{tab:modulispaces}
\end{table}
\endgroup

In the subsequent sections, we discuss the symplectic modular transformations of matter fields associated with the complex structure moduli and K\"ahler moduli. 
The flavor structure of matter fields can be read off from 
the symplectic structure of moduli Yukawa couplings.

\subsection{Symplectic modular flavor symmetry of matter fields}
\label{sec:modularflavor}

In general, it is difficult to derive the flavor structure of matter fields in smooth CY compactifications due to the lack of analytical expression of the CY metric. 
Hence, we concentrate on the heterotic string theory with standard embedding to overcome this problem. 
The matter fields $27_a$ and $\widebar{27}_i$ under the $E_6$ gauge group are in one-to-one correspondence with the K\"ahler moduli and the complex structure moduli, respectively. It indicates that the flavor structure of matter fields is directly related to the symplectic structure in the moduli spaces. 
In the following discussions, we assume that the matter zero-modes appear from either the complex structure moduli sector or the K\"ahler moduli sector since the net number of zero-modes is determined by
\begin{align}
    \chi = \frac{1}{2}|h^{2,1}- h^{1,1}|,
\end{align}
through the Atiyah-Singer index theorem. 
We will focus on the matter fields associated with either the complex structure moduli or the K\"ahler moduli 
in order to reveal the relation between the flavor symmetry of matter fields and symplectic modular symmetry appearing in the moduli effective action. 

It was known that the matter K\"ahler metric is of the form\footnote{The overall factor $e^{\pm\frac{1}{3}(K_{\rm cs}-K_{\rm ks})}$ takes the different form on toroidal orbifolds in the example of section \ref{sec:3} due to the enlarged symmetry\cite{Dixon:1989fj}.}
\begin{align}
    K^{(27)}_{a\widebar{b}} &= e^{\frac{1}{3}(K_{\rm cs}-K_{\rm ks})}(K_{\rm ks})_{a\widebar{b}},
    \nonumber\\
   K^{(\widebar{27})}_{i\widebar{j}} &= e^{-\frac{1}{3}(K_{\rm cs}-K_{\rm ks})}(K_{\rm cs})_{i\widebar{j}},
   \label{eq:KahlerMetric}
\end{align}
where $(K_{\rm cs})_{i\widebar{j}}=\partial_i\partial_{\bar{j}}K_{\rm cs}$, $(K_{\rm ks})_{a\widebar{b}}=\partial_a\partial_{\bar{b}}K_{\rm ks}$, $A^i$ and $A^a$ denote the matter chiral superfields, respectively.\footnote{In the following, we call matter chiral superfields as matters for simplicity.}  
At the moment, we study the relation between $\widebar{27}_i$ matter fields and the complex structure moduli $u^i$ for concreteness. 
The K\"ahler potential of the complex structure moduli in the large complex structure regime
\begin{align}
    K_{\rm cs} &= - \ln \biggl[-i \int_{\cal M} \Omega \wedge \widebar{\Omega}\biggl] 
    = - \ln \biggl[-i |X^0|^2 \frac{\kappa_{ijk}}{6} (u^i - \widebar{u}^i)(u^j - \widebar{u}^j)(u^k - \widebar{u}^k)\biggl],
\end{align}
indicates that the matter K\"ahler metric obeys 
\begin{align}
            K^{(\widebar{27})}_{i\widebar{j}} \propto |X^0|^{2/3}(K_{\rm cs})_{i\widebar{j}},
            \label{eq:X0matter}
\end{align}
up to the factor relevant to the K\"ahler moduli. 
Hence, symplectic transformations of $\widebar{27}_i$ matter fields are almost the same with those of the moduli fields $u^i$, up to the $X^0$ factor. 
Indeed, the matter K\"ahler metric transforms under the symplectic modular transformations as
\begin{align}
            K^{(\widebar{27})}_{i\widebar{j}} \rightarrow |\widetilde{X}^0|^{2/3}
             (K_{\rm cs})_{\widetilde{l}\widetilde{\widebar{m}}}
            \pdv{\widetilde{u}^i}{u^l}\pdv{\widetilde{\widebar{u}}^j}{\widebar{u}^m}
\end{align}
up to the factor $e^{K_{\rm ks}/3}$. 
It turns out that the K\"ahler potential of $\widebar{27}_i$ matter fields has a modular invariance 
under the following symplectic transformation of matter fields:
\begin{align}
   A^i \rightarrow \widetilde{A}^i =  (\widetilde{X}^0)^{1/3} \pdv{\widetilde{u}^i}{u^l} A^l
                                                =  \frac{\partial \widetilde{X}^i}{\partial X^j}A^j,
    \label{eq:matterTrans}
\end{align}
where we use Eq. (\ref{eq:tildeFijk3}). 
Since the matrix
\begin{align}
   \frac{\partial \widetilde{X}^i}{\partial X^j} =  (c{\cal F}+d)^{i}_{j}
\end{align}
is an element of the subgroup of $Sp(2h^{2,1}+2,\mathbb{Z})$, the flavor structure of matter fields is governed by the symplectic structure. 
In what follows, we adopt the gauge $X^0=1$ unless we specify it. 
As will be discussed later, 
the above modular transformation is indeed consistent with the K\"ahler invariance of the action 
by checking the K\"ahler invariant quantity.

On the other hand, the matter superpotential takes the form
\begin{align}
    W &=  F_{ijk}A^iA^jA^k,
    \label{eq:Wmatter}
\end{align}
where the matter Yukawa couplings are the same as the Yukawa couplings of the moduli fields. 
As we derived before, the $Sp(2h^{2,1}+2,\mathbb{Z})$ modular transformations of their Yukawa couplings 
are provided by 
\begin{align}
        \widetilde{F}_{ijk} &= \widetilde{X}^0\pdv{X^l}{\widetilde{X}^i}\pdv{X^m}{\widetilde{X}^j}\pdv{X^n}{\widetilde{X}^k}\,F_{lmn}.
        \end{align}
Hence, we arrive at the symplectic modular transformation of the K\"ahler potential and the superpotential:
\begin{align}
    K_{\rm cs} &\rightarrow K_{\rm cs} -\ln |\widetilde{X}^0|^2,
    \quad
    W \rightarrow \widetilde{X}^0W,
\end{align}
from which the K\"ahler invariant quantity
\begin{align}
    e^K|W|^2 
\end{align}
is invariant under the symplectic modular transformation.

We conclude that the flavor symmetry of matter fields $\widebar{27}_i$, $G_{\rm flavor}$, belongs to the $Sp(2h^{2,1}+2, \mathbb{Z})$ modular symmetry,
\begin{align}
 G_{\rm flavor}\subset Sp (2h^{2,1}+2, \mathbb{Z}),
\end{align}
and the flavor structure is governed by the symplectic structure of the 
complex structure moduli. The same statements also hold for the $27_a$ matters associated with the K\"ahler 
moduli, whose prepotential is given in Eq. (\ref{eq:G}) in the large volume regime. 
The flavor structure of $27_a$ matters is determined by the holomorphic Yukawa coupling $\kappa_{abc}$. 
For that reason, we call the flavor symmetry of matter fields the {\it symplectic modular flavor symmetry}. 
The symplectic transformations of matter zero-modes, moduli, and holomorphic Yukawa couplings are summarized in Table \ref{tab:modularflavor}. 
It is a natural generalization of $Sp(2,\mathbb{Z})\simeq SL(2,\mathbb{Z})$ discussed in the toroidal background \cite{Feruglio:2017spp}. 
Since it is difficult to derive the flavor symmetries in a background-independent way, we explicitly demonstrate the existence of symplectic modular flavor symmetries on toroidal orbifolds with and without resolutions and some classes of CY threefolds in Sec. \ref{sec:3}. 

\begingroup
\renewcommand{\arraystretch}{1.5}
\begin{table}[H]
   \begin{center}
   \scalebox{0.9}{
    \begin{tabular}{ccc} \toprule
       & $\widebar{27}$ matter fields & $27$ matter fields \\ 
       & (Complex structure) & (K\"ahler structure) \\ \midrule
      Matter fields & $A^i$ & $A^a$\\ 
      Moduli fields & $u^i$ & $t^a$\\ 
      Superpotential & $F_{ijk}A^iA^jA^k$ & $G_{abc}A^aA^bA^c$\\ 
      Symplectic transformations & $A^i \rightarrow \widetilde{A}^i =\frac{\partial \widetilde{X}^i}{\partial X^j} A^j$ & $A^a \rightarrow \widetilde{A}^a = \frac{\partial \widetilde{Y}^a}{\partial Y^b} A^b$\\ 
                                 & $u^i \rightarrow \widetilde{u}^i =(\widetilde{X}^0)^{-1} \frac{\partial \widetilde{X}^i}{\partial X^j} u^j$ 
                                 & $t^a \rightarrow \widetilde{t}^a =(\widetilde{Y}^0)^{-1} \frac{\partial \widetilde{Y}^a}{\partial Y^b} t^b$\\ 
                                 & $F_{ijk}\rightarrow \widetilde{F}_{ijk} = \widetilde{X}^0\pdv{X^l}{\widetilde{X}^i}\pdv{X^m}{\widetilde{X}^j}\pdv{X^n}{\widetilde{X}^k}F_{lmn}$
                                 & $G_{abc}\rightarrow \widetilde{G}_{abc} = \widetilde{Y}^0\pdv{Y^d}{\widetilde{Y}^a}\pdv{Y^e}{\widetilde{Y}^b}\pdv{Y^f}{\widetilde{Y}^c}G_{def}$\\
                                 & $\frac{\partial \widetilde{X}^i}{\partial X^j} =(c{\cal F}+d)^{i}_{j}$
                                 & $\frac{\partial \widetilde{Y}^a}{\partial Y^b} =(g{\cal G}+h)^{a}_{b}$\\
      Flavor symmetry & $G_{\rm flavor} \subset Sp(2h^{2,1}+2,\mathbb{Z})$ & $G_{\rm flavor} \subset Sp(2h^{1,1}+2,\mathbb{Z})$\\ 
      \bottomrule
    \end{tabular}
    }
  \end{center}
  \caption{Symplectic modular flavor symmetries.}
  \label{tab:modularflavor}
\end{table}
\endgroup

\subsection{$U(1)_R$ symmetry}
\label{sec:U(1)R}

In addition to the symplectic modular flavor symmetry, there exists a $U(1)_R$ symmetry 
in the 4D ${\cal N}=1$ supersymmetric effective action. 
From the cubic matter superpotential (\ref{eq:Wmatter}), one can assign the $R$-charge 2/3 
for the matter fields in a conventional way due to the fact that the superpotential has $R$-charge 2. 
Here, we examine the $\widebar{27}_i$ matter fields, but $27_a$ 
matter fields also have $R$-charge 2/3 due to the same cubic superpotential. 
In this section, we discuss the relation between the symplectic modular symmetry and 
$U(1)_R$ symmetry. 

Let us consider the symplectic transformations in the basis where the Yukawa couplings do not have the factor $\tilde{X}^0$,
\begin{align}
   A^i &\rightarrow \widetilde{A}^i =  (\widetilde{X}^0)^{1/3} \pdv{\widetilde{u}^i}{u^l} A^l,
   \nonumber\\
        F_{ijk} &\rightarrow \widetilde{F}_{ijk} =
        \pdv{u^l}{\widetilde{u}^i}\pdv{u^m}{\widetilde{u}^j}\pdv{u^n}{\widetilde{u}^k}F_{lmn}.
\end{align}
Since the $U(1)_R$ symmetry is not the flavor symmetry, the $(\widetilde{X}^0)^{1/3}$ factor 
appearing in the symplectic transformation of matter fields would play a role of $R$-symmetric transformation. 
In this respect, let us set 
\begin{align}
    \widetilde{X}^0 = e^{2i\alpha},
\end{align}
with $\alpha$ being a constant relevant to the $R$-symmetric transformation\cite{Billo:1995yq}. 
This assignment is consistent with the symplectic modular symmetry and 
it gives rise to the $R$-charge 2 for the moduli fields. 
Indeed, the $R$-symmetric transformations for the period integrals in the large complex 
structure regime correspond to
\begin{align}
    \begin{pmatrix}
       X^I\\
       {\cal F}_I\\
    \end{pmatrix}
\rightarrow
    \begin{pmatrix}
       \widetilde{X}^I\\
       \widetilde{{\cal F}}_I\\
    \end{pmatrix}
=\widetilde{X}^0
    \begin{pmatrix}
       X^I\\
       {\cal F}_I\\
    \end{pmatrix}
    ,
\end{align}
with 
\begin{align}
    \{X^0,X^i, {\cal F}_0, {\cal F}_i\} &=\left\{ 1, u^i, -\frac{1}{3!}\kappa_{ijk}u^iu^ju^k, \frac{1}{2}\kappa_{ijk}u^ju^k\right\},
   \nonumber\\
    \{\widetilde{X}^0,\widetilde{X}^i, \widetilde{{\cal F}}_0, \widetilde{{\cal F}}_i\} &=\widetilde{X}^0\left\{ 1, \widetilde{u}^i, -\frac{1}{3!}\kappa_{ijk}\widetilde{u}^i\widetilde{u}^j \widetilde{u}^k, \frac{1}{2}\kappa_{ijk}\widetilde{u}^j \widetilde{u}^k\right\}.
\end{align}
As stated in Ref.~\cite{Witten:1985xc}, $R$-symmetric transformation corresponds to the 
holomorphic transformation $f$ mapping the holomorphic three-form of CY threefolds $\Omega$ into a multiple of itself,
\begin{align}
    f\cdot \Omega = e^{i\gamma} \Omega,
\end{align}
with $\gamma$ being the constant. 
Recalling that the $\Omega$ is constructed by a covariantly constant positive-chirality spinor $\eta$, namely $\Omega_{ijk}=\eta^T \Gamma_{ijk}\eta$ with three products of Gamma matrix $\Gamma_{ijk}$, 
the holomorphic mapping $f$ rotates the $\eta$, 
\begin{align}
    f\cdot \eta = e^{i\gamma/2}\eta,
\end{align}
namely the 4D ${\cal N}=1$ supersymmetry by a phase of $e^{i\gamma/2}$ for positive chirality spinor. The negative chirality spinor is rotated by an amount of $e^{-i\gamma/2}$. 
In this way, the phase shift of $\Omega$ gives rise to the $R$-symmetry in 4D ${\cal N}=1$ supersymmetric theory. 
In our setup, our finding $R$-symmetric transformation $\widetilde{X}^0=e^{2i\alpha}$ corresponds to $\gamma=2\alpha$. 
Note that the existence of $U(1)_R$ symmetry is related to the discrete $\mathbb{Z}_{3}^R$ $R$-symmetry appearing in the matter superpotential,
\begin{align}
   W = F_{ijk}A^iA^jA^k, 
\end{align}
by assigning the charge 1 for all the matter fields. 
When we focus on the overall complex structure modulus $u^1=u^2=\cdots =u^{h^{2,1}}$, 
the action is invariant under the overall $SL(2,\mathbb{Z})_{\rm overall}$ symmetry. 
From the fact that the superpotential has modular weight -3 under $SL(2,\mathbb{Z})_{\rm overall}$, the matter fields have charge -1. That would be related to the origin of charge 1 under the discrete $\mathbb{Z}_{3}^R$ $R$-symmetry \footnote{The relations between 
$R$-symmetry and the modular symmetry, in particular, their anomalies were studied in Refs. \cite{Araki:2007ss,Araki:2008ek}.}. 
Similar statements also hold for the K\"ahler moduli space, by identifying $Y^0$ with the $R$-symmetric transformations of $27_a$ matter fields. 
Hence, $R$-symmetric transformations would be related to the phase rotation of the axio-dilaton. 
It is notable that the K\"ahler form consists of Killing spinors, $J_{m\bar{n}}= i \eta^\dagger \Gamma_{m\bar{n}}\eta$ with two products of Gamma matrix $\Gamma_{m\bar{n}}$ in the complex basis $z^m$ with $m=1,2,3$, but it does not change under the $R$-transformation. 

As a result, $U(1)_R$ symmetry is also unified in the context of symplectic modular 
symmetry, i.e., 
\begin{align}
    Sp (2h+2, \mathbb{C}) \supset Sp (2h+2, \mathbb{Z}) \times U(1)_R,
\end{align}
with $h=h^{2,1}$ or $h^{1,1}$.

\subsection{Unification of symplectic modular flavor, CP, and $R$-symmetries}
\label{sec:unification}

Following Refs. \cite{Strominger:1985it,Dine:1992ya,Choi:1992xp}, one can identify the 4D CP symmetry with the simultaneous transformation of 4D parity and the 6D orientation reversing. 
They belong to the 10D proper Lorentz transformation. 
We recall that the 10D Majorana-Weyl spinor $16$ decomposes under the 10D Lorentz symmetry $SO(1,9)=SO(1,3)\times SO(6)$,
\begin{align}
    16 = (2_L, 4_+) \oplus (2_R, \widebar{4}_-)
\end{align}
where $2_L$ and $2_R$ denote the left- and right-handed spinors of $SL(2,\mathbb{C})$, and $4_+, \widebar{4}_-$ represent the positive and negative chirality spinors of $SU(4) \simeq SO(6)$, respectively. 
Hence, the 4D parity and 6D orientation reversing exchanges $(2_L, 4_+)$ into $(2_R, \widebar{4}_-)$ which is in one-to-one correspondence with the 4D CP transformation of $E_6$ matter fields: $27\rightarrow \widebar{27}$. 
It suggests that the 4D CP transformation is identified with the geometrical symmetry of the internal manifold. 

The 6D orientation reversing changes the sign of the volume form of CY threefolds. 
In the local coordinate of CY threefolds $z^m$ with $m=1,2,3$, 
the following transformation
\begin{align}
    z^m \rightarrow - \widebar{z}^m
\label{eq:6Dorient}
\end{align}
changes the sign of the volume form.\footnote{The minus sign is a matter of convention.} 
To see the orientation reversing of the volume form, we examine the transformation of the holomorphic three-form $\Omega$ and K\"ahler form $J$. 
Eq. (\ref{eq:6Dorient}) gives rise to the $\mathbb{Z}_2^{\rm CP}$ transformation:
\begin{align}
\begin{split}
    \Omega \rightarrow - \widebar{\Omega},
    \\
    J_c \rightarrow \widebar{J_c},
\end{split}
    \label{eq:OmegaTrans}
\end{align}
where the anti-holomorphic transformation of $\Omega$ is consistent with its local expansion $\Omega = dz^1\wedge dz^2 \wedge dz^3$. 
On CY threefolds, the 6D volume form $dV$ is expressed by the $\Omega$,
\begin{align}
    i \Omega \wedge \widebar{\Omega} = |\Omega|^2 dV,
\end{align}
thereby the transformation (\ref{eq:OmegaTrans}) changes the sign of the volume form, namely the 6D orientation reversing. 

We comment on the orientation reversing of the K\"ahler form $J_c$ which is locally given by $J_c = B+ iJ = (b_{m\bar{n}}+ig_{m\bar{n}}) dz^m \wedge d\widebar{z}^{\bar{n}}$, respectively. 
Recall that the real part of complexified K\"ahler form ($b_{m\bar{n}}$) originates from the Kalb-Ramond field, leading to the pseudo scalars (axions) in the 4D effective action through the hodge duality. 
Since the axions are 4D CP-odd fields through the 4D Chern-Simons couplings, 
$b_{m\bar{n}}$ is required to be transformed as $b_{m\bar{n}}\rightarrow - b_{m\bar{n}}$. 
By contrast, the imaginary part of complexified K\"ahler form ($g_{m\bar{n}}$) is a CP-even field. 
Hence, we arrive at the CP transformation of the complexified K\"ahler form as in Eq. (\ref{eq:OmegaTrans}).

As discussed in Ref. \cite{Strominger:1985ku}, the anti-holomorphic transformation of the holomorphic three-form induces the anti-holomorphic transformation of the complex structure moduli, and the CP-transformation of the complexified K\"ahler 
form also induces the anti-holomorphic transformation of the K\"ahler moduli due to the fact that $b_{m\bar{n}}$ and $g_{m\bar{n}}$ are CP-odd and -even fields, respectively. 
Then, we arrive at the CP transformation of the moduli fields:
\begin{align}
u^i \rightarrow -\widebar{u}^i,
\quad
t^a \rightarrow -\widebar{t}^a,
\end{align}
where we take the convention with ${\rm Im}(u^i)>0$ and ${\rm Im}(t^a)>0$. 
Furthermore, since the real part of the axio-dilaton $Y^0$  originating from the Kalb-Ramond six-form is the universal axion, 
the CP transformation gives rise to
\begin{align}
Y^0 \rightarrow -\widebar{Y}^0.
\end{align}
It is trivial to check that these anti-holomorphic transformations are not the element of symplectic modular 
groups, but they belong to the generalized symplectic modular group $GSp(2g+2,\mathbb{Z})\simeq Sp(2g+2,\mathbb{Z}) \rtimes \mathbb{Z}_2^{\rm CP}$ with $g=h^{2,1},h^{1,1}$. 
Indeed, the $\mathbb{Z}_2^{\rm CP}$ transformation acts on the projective coordinates in the complex structure and the K\"ahler structure moduli spaces\footnote{Here, we adopt the minus sign for the CP transformation in the complex structure 
moduli space, although the positive sign is allowed.}
\begin{align}
    &\begin{pmatrix}
       X^0\\
       X^i\\
       {\cal F}_0\\
       {\cal F}_i\\
    \end{pmatrix}
    \rightarrow
     -
     \begin{pmatrix}
       \widebar{X}^0\\
       -\widebar{X}^i\\
       -\widebar{{\cal F}}_0\\
       \widebar{{\cal F}}_i\\
    \end{pmatrix}   
      = {\cal CP}
    \begin{pmatrix}
       X^0\\
       X^i\\
       {\cal F}_0\\
       {\cal F}_i\\
    \end{pmatrix}
    ,\quad
    \begin{pmatrix}
       Y^0\\
       Y^a\\
       {\cal G}_0\\
       {\cal G}_a\\
    \end{pmatrix}
    \rightarrow
     -
     \begin{pmatrix}
       \widebar{Y}^0\\
       -\widebar{Y}^a\\
       -\widebar{{\cal G}}_0\\
       \widebar{{\cal G}}_a\\
    \end{pmatrix}   
      ={\cal CP}
    \begin{pmatrix}
       Y^0\\
       Y^a\\
       {\cal G}_0\\
       {\cal G}_a\\
    \end{pmatrix}
\label{eq:CPtrf}
    ,
\end{align}
with
\begin{align}
{\cal CP}=
     \begin{pmatrix}
       -1 & 0 & 0 & 0\\
       0 & \bf{1} & 0 & 0\\
       0 & 0 & 1 & 0\\
       0 & 0 & 0 & -\bf{1}\\
    \end{pmatrix}
    \notin Sp(2h^{2,1}+2, \mathbb{Z}),Sp(2h^{1,1}+2, \mathbb{Z}).
\label{eq:CPdef}
\end{align}
Here, we use the shorthand notation for the identity matrix $\bf{1}$, which is rank $h^{2,1}$ for the complex structure moduli and $h^{1,1}$ for the K\"ahler moduli, respectively.
Hence, they do not belong to the element of $Sp(2h^{2,1}+2, \mathbb{Z})$ and $Sp(2h^{1,1}+2, \mathbb{Z})$, 
due to the fact that the CP-transformation does not have the symplectic structure ${\cal CP}^T
 \cdot \Sigma \cdot 
{\cal CP}
=\Sigma^T \neq \Sigma$. 
Rather, the CP transformation is regarded as an outer automorphism of the symplectic modular group, i.e., ${\cal CP}\cdot \gamma \cdot {\cal CP}^{-1}$ with $\gamma$ being the transformation of the $Sp(2h^{2,1}+2, \mathbb{Z})$ or $Sp(2h^{1,1}+2, \mathbb{Z})$ modular group\cite{Ishiguro:2020nuf}\footnote{Outer automorphisms of flavor symmetries were studied as 
generalized CP symmetries  \cite{Feruglio:2012cw,Holthausen:2012dk,Chen:2014tpa}.}.

As a result, the symplectic modular group together with the CP transformation is enlarged to the generalized symplectic modular group. In particular, the flavor symmetry belonging to the symplectic modular symmetry is also enhanced to the 
non-Abelian symmetry,
\begin{align}
    GSp(2h^{2,1}+2,\mathbb{Z}) &\supset G_{\rm flavor} \rtimes \mathbb{Z}_2^{\rm CP},
    \nonumber\\
    GSp(2h^{1,1}+2,\mathbb{Z}) &\supset G_{\rm flavor} \rtimes \mathbb{Z}_2^{\rm CP},    
\end{align}
as exemplified in the next section. The $U(1)_R$ symmetry further enhances them into
\begin{align}
    GSp(2h^{2,1}+2,\mathbb{C}) &\supset G_{\rm flavor} \rtimes \mathbb{Z}_2^{\rm CP} \times U(1)_R,
    \nonumber\\
    GSp(2h^{1,1}+2,\mathbb{C}) &\supset G_{\rm flavor} \rtimes \mathbb{Z}_2^{\rm CP} \times U(1)_R.    
\end{align}

In the following analysis, we explicitly demonstrate the symplectic modular flavor symmetry on several backgrounds. 
This approach would be deserved to be the 
ultra-violet completion of the phenomenological model buildings in the bottom-up approach, where the symplectic modular flavor symmetry is used to explain the flavor structure of quarks and leptons.

\section{Examples}
\label{sec:3}

In this section, we show the existence of generalized modular symmetry involving the flavor and CP symmetries 
on several backgrounds. 
Starting from the simplest 6D toroidal orbifolds in Sec. \ref{sec:orb}, we deal with several blown-up toroidal orbifolds, 
where the size of blow-up radii is considered small enough in Secs. \ref{sec:smallblowZ3} and \ref{sec:smallblowZ3Z3} 
and large enough in Sec. \ref{sec:largeblowZ3Z3}. 
The three-parameter examples of CY threefolds are analyzed in Sec. \ref{sec:CY}.

\subsection{6D toroidal orbifolds}
\label{sec:orb}

We begin with toroidal orbifolds with an emphasis on the flavor symmetry of untwisted modes. The twisted (blow-up) modes are analyzed in detail in the next sections. 
The simplest but non-trivial examples realizing the flavor symmetries of untwisted modes are based on $T^6/\mathbb{Z}_N$ and $T^6/(\mathbb{Z}_N\times \mathbb{Z}_M)$ with $h^{1,1}_{\rm untw}=3$, whose prepotential is given by\footnote{The following discussions are applicable to the sector of $h^{2,1}=3$ complex structure moduli sector as well.}
\begin{align}
    G = t^1 t^2 t^3,
\end{align}
leading to the K\"ahler potential\cite{Dixon:1989fj}
\begin{align}
    K_{\rm ks} = -\ln \biggl[i(t^1 - \widebar{t}^1)(t^2 - \widebar{t}^2)(t^3 - \widebar{t}^3) -\sum_a |A^a|^2\biggl] - \ln |Y^0|^2.
    \label{eq:Korbifold}
\end{align}
Here, we include $Y^0$ for the later purpose, and the matter fields are assumed to be expanded around $|A^a| \ll 1$. 
Correspondingly, the matter superpotential is of the form
\begin{align}
W = Y^0 A^1 A^2 A^3.
\label{eq:WA4}
\end{align}

Obviously,  there appear the $\Pi_{a=1}^3 SL(2,\mathbb{Z})_a$ symmetries generated by
\begin{align}
    S^{(a)} : t^a \rightarrow -1/t^a,
    \quad
    T^{(a)} : t^a \rightarrow t^a+1,    
\end{align}
for $\Pi_{a=1}^3 SL(2,\mathbb{Z})_a$. They originate from factorizable three tori: $T^6=(T^2)^3$. 
In addition,  
we find that there exists a $S_4$ symmetry by assigning $\{ t^1, t^2, t^3\}$ for $S_4$ triplet and $Y^0$ 
for $S_4$ singlet. 
Although the $S_4$ group has two triplets $\{\mathbf{3},\mathbf{3}^\prime\}$ and two singlets 
$\{\mathbf{1},\mathbf{1}^\prime\}$, the superpotential is invariant under two cases:
(i) $\{ t^1, t^2, t^3\}$ for $\mathbf{3}$ and $Y^0$ for $\mathbf{1}$, (ii) 
$\{ t^1, t^2, t^3\}$ for $\mathbf{3}^\prime$ and $Y^0$ for $\mathbf{1}^\prime$. 
(See for non-Abelian discrete symmetries and notation of representation, Refs.~\cite{Ishimori:2010au,Ishimori:2012zz}. )
When we take the gauge $Y^0=1$, namely $S_4$ trivial singlet $\mathbf{1}$, 
the transformation of the $S_4$ triplet $\{ t^1, t^2, t^3\}$ is given by
\begin{align}
    P &: t^1 \rightarrow t^3,\quad t^2 \rightarrow t^1,\quad t^3 \rightarrow t^2,  
    \nonumber\\
    Q &: t^1 \rightarrow -t^1,\quad t^2 \rightarrow -t^3,\quad t^3 \rightarrow t^2,        
    \label{eq:PQ}
\end{align}
where $P$ corresponds to $S_3$ permutation group of three $SL(2,\mathbb{Z})_a$ \cite{Shevitz:1990pc}. 
In the context of heterotic string theory with standard embedding, 
these symmetries are identified with the flavor symmetries of matter fields. 
Indeed, the matter fields $\{ A^1, A^2, A^3\}$  are also identified with the $S_4$ triplet representation 
under which the effective action of matter fields is invariant. 

Our purpose in this section is to check whether these symmetries together with the CP transformation can be embedded into the $GSp(8, \mathbb{Z})$ generalized symplectic modular symmetry. 
The symplectic structure is understood from the symplectic transformation of the period vector
\begin{align}
    \begin{pmatrix}
       Y^A\\
       {\cal G}_A\\
    \end{pmatrix}
    =
    \begin{pmatrix}
    Y^0\\
    Y^1\\
    Y^2\\
    Y^3\\
    {\cal G}_0\\
    {\cal G}_1\\
    {\cal G}_2\\
    {\cal G}_3\\
    \end{pmatrix}
    =
    \begin{pmatrix}
    Y^0\\
    Y^1\\
    Y^2\\
    Y^3\\
    -Y^1Y^2Y^3/(Y^0)^2\\
    Y^2Y^3/Y^0\\
    Y^1Y^3/Y^0\\
    Y^1Y^2/Y^0\\
    \end{pmatrix}
    =Y^0
    \begin{pmatrix}
    1\\
    t^1\\
    t^2\\
    t^3\\
    -t^1 t^2 t^3\\
    t^2t^3\\
    t^1t^3\\
    t^1t^2\\
    \end{pmatrix}  
,
\end{align}
leading to the K\"ahler potential (\ref{eq:Korbifold}). 
Since the period vector is a fundamental representation of $Sp(8,\mathbb{Z})$, 
the symplectic transformation is represented by
\begin{align}
    \begin{pmatrix}
       Y^A\\
       {\cal G}_A\\
    \end{pmatrix}
    =Y^0
    \begin{pmatrix}
    1\\
    t^1\\
    t^2\\
    t^3\\
    -t^1 t^2 t^3\\
    t^2t^3\\
    t^1t^3\\
    t^1t^2\\
    \end{pmatrix}  
\rightarrow
    \widetilde{Y}^0
    \begin{pmatrix}
    1\\
    \widetilde{t}^1\\
    \widetilde{t}^2\\
    \widetilde{t}^3\\
    -\widetilde{t}^1 \widetilde{t}^2 \widetilde{t}^3\\
    \widetilde{t}^2\widetilde{t}^3\\
    \widetilde{t}^1\widetilde{t}^3\\
    \widetilde{t}^1\widetilde{t}^2\\
    \end{pmatrix}  
= Y^0\times {\cal R}
    \begin{pmatrix}
    1\\
    t^1\\
    t^2\\
    t^3\\
    -t^1 t^2 t^3\\
    t^2t^3\\
    t^1t^3\\
    t^1t^2\\
    \end{pmatrix} 
    ,
\end{align}
with ${\cal R}\subset Sp(8,\mathbb{Z})$.

It turns out that $\Pi_{a=1}^3 SL(2,\mathbb{Z})_a$ and $S_4$ symmetries are indeed subgroups of $Sp(8,\mathbb{Z})$. 
For instance, the generators of $SL(2,\mathbb{Z})_1$ and  $S_4$ are chosen as\footnote{It is straightforward to obtain the generators of other $SL(2,\mathbb{Z})$ by flipping the corresponding 
moduli fields.}
 \begin{align}
 \begin{split}
S_1&=
\begin{pmatrix}
0 & -1 & 0 & 0 & 0 & 0 & 0 & 0\\
1 & 0 & 0 & 0 & 0 & 0 & 0 & 0\\
0 & 0 & 0 & 0 & 0 & 0 & 0 & -1\\
0 & 0 & 0 & 0 & 0 & 0 & -1 & 0\\
0 & 0 & 0 & 0 & 0 & -1 & 0 & 0\\
0 & 0 & 0 & 0 & 1 & 0 & 0 & 0\\
0 & 0 & 0 & 1 & 0 & 0 & 0 & 0\\
0 & 0 & 1 & 0 & 0 & 0 & 0 & 0
\end{pmatrix}
, \quad 
T_1=
\begin{pmatrix}
1 & 0 & 0 & 0 & 0 & 0 & 0 & 0\\
1 & 1 & 0 & 0 & 0 & 0 & 0 & 0\\
0 & 0 & 1 & 0 & 0 & 0 & 0 & 0\\
0 & 0 & 0 & 1 & 0 & 0 & 0 & 0\\
0 & 0 & 0 & 0 & 1 & -1 & 0 & 0\\
0 & 0 & 0 & 0 & 0 & 1 & 0 & 0\\
0 & 0 & 0 & 1 & 0 & 0 & 1 & 0\\
0 & 0 & 1 & 0 & 0 & 0 & 0 & 1
\end{pmatrix}
,
\\
P&=
\begin{pmatrix}
1 & 0 & 0 & 0 & 0 & 0 & 0 & 0\\
0 & 0 & 0 & 1 & 0 & 0 & 0 & 0\\
0 & 1 & 0 & 0 & 0 & 0 & 0 & 0\\
0 & 0 & 1 & 0 & 0 & 0 & 0 & 0\\
0 & 0 & 0 & 0 & 1 & 0 & 0 & 0\\
0 & 0 & 0 & 0 & 0 & 0 & 0 & 1\\
0 & 0 & 0 & 0 & 0 & 1 & 0 & 0\\
0 & 0 & 0 & 0 & 0 & 0 & 1 & 0
\end{pmatrix}
,
\quad
Q=
\begin{pmatrix}
1 & 0 & 0 & 0 & 0 & 0 & 0 & 0\\
0 & -1 & 0 & 0 & 0 & 0 & 0 & 0\\
0 & 0 & 0 & -1 & 0 & 0 & 0 & 0\\
0 & 0 & 1 & 0 & 0 & 0 & 0 & 0\\
0 & 0 & 0 & 0 & 1 & 0 & 0 & 0\\
0 & 0 & 0 & 0 & 0 & -1 & 0 & 0\\
0 & 0 & 0 & 0 & 0 & 0 & 0 & -1\\
0 & 0 & 0 & 0 & 0 & 0 & 1 & 0
\end{pmatrix}
,
\end{split}
\end{align}
where the fundamental representation of $Sp(8,\mathbb{Z})$ is split into 
\begin{align}
\mathbf{8} = \mathbf{1}+ \mathbf{3} + \mathbf{1} + \mathbf{3},
\end{align}
under $S_4$ symmetry. 
We have focused on the transformation of moduli fields, but the $S_4$ flavor symmetry of matter fields is 
also realized by $P$ and $Q$. 
Indeed, the submatrices of $P$ and $Q$ acting on $(t^1,t^2,t^3)^T$ 
\begin{align}
\begin{pmatrix}
0 & 0 & 1 \\
1 & 0 & 0 \\
0 & 1 & 0 \\
\end{pmatrix}
,\quad
\begin{pmatrix}
-1 & 0 & 0 \\
0 & 0 & -1 \\
0 & 1 & 0 \\
\end{pmatrix}
,
\label{eq:A4Trf}
 \end{align}
correspond to the transformations for untwisted matters $\{A^1,A^2,A^3\}$ with the $S_4$ triplet. 
Note that the holomorphic Yukawa coupling is invariant under this $S_4$ symmetry.

Let us discuss the origin of the $S_4$ symmetry in the effective action. 
We recall that imaginary parts of moduli ${\rm Im}(t^a)$ behave as a triplet representation under $SO(3)$ 
group due to the fact that the volume form of factorizable toroidal background ${\cal V}$ is given by 
${\cal V} \simeq \otimes_{a=1}^3{\rm Im}(t^a)dx^a\wedge dy^a$. 
Since the $SO(3)$ triplet representation corresponds to the $S_4$ triplet representation\footnote{See, e.g., Ref. \cite{Rachlin:2017rvm} and references therein.}
, $S_4$ symmetry exists in the moduli sector $\{{\rm Im}(t^1), {\rm Im}(t^2), {\rm Im}(t^3)\}$. 
On the complexified K\"ahler form, the K\"ahler moduli consist of volume moduli ${\rm Im}(t^a)$ and the 
K\"ahler axions ${\rm Re}(t^a)$. This fact enhances $SO(3)$ into $SU(3)$ symmetry. 
On the other hand, matter fields are originated from 10D gauge bosons and gauginos. 
Since 10D ${\cal N}=1$ supersymmetry gives rise to 4D ${\cal N}=4$ supersymmetry equipping $SU(4)_R$ symmetry 
on $T^6$ background without orbifolding, 
the fundamental representation of $SU(4)_R$ symmetry can be split into 
\begin{align}
\mathbf{4} = \mathbf{1}+ \mathbf{3},
\end{align}
under the $SU(3)_R$ symmetry.  
The matter fields $\{A^a\}$ of our interest have triplet representation of $SU(3)_R$ symmetry, corresponding to the 
$S_4$ triplet representation.  
Indeed, the $SU(3)_R$ symmetry is embedded into $Sp(8,\mathbb{C})$ as follows:
 \begin{align}
 \begin{split}
R &=
\begin{pmatrix}
1 & 0 & 0 & 0 & 0 & 0 & 0 & 0\\
0 & a_1 & a_2 & a_3 & 0 & 0 & 0 & 0\\
0 & a_4 & a_5 & a_6 & 0 & 0 & 0 & 0\\
0 & a_7 & a_8 & a_9 & 0 & 0 & 0 & 0\\
0 & 0 & 0 & 0 & 1 & 0 & 0 & 0\\
0 & 0 & 0 & 0 & 0 &  a_1 & a_2 & a_3\\
0 & 0 & 0 & 0 & 0 & a_4 & a_5 & a_6 \\
0 & 0 & 0 & 0 & 0 & a_7 & a_8 & a_9
\end{pmatrix}
,\quad
{\rm with}
\begin{pmatrix}
a_1 & a_2 & a_3\\
a_4 & a_5 & a_6\\
a_7 & a_8 & a_9\\
\end{pmatrix}
\in SU(3)_R,
 \end{split}
 \end{align}
where we used that $t^1t^2t^3$ ($A^1A^2A^3$) is a $SU(3)_R$ singlet.

In this way, $S_4$ flavor symmetry naturally lies in the symplectic modular flavor symmetry. Together with the CP transformation as discussed in Sec. \ref{sec:unification}, the symplectic modular symmetry is generalized to the generalized symplectic modular symmetry. 
It turns out that the $S_4$ flavor symmetry is enhanced to 
 \begin{align}
     S_4 \rtimes \mathbb{Z}_2^{\rm CP},
 \end{align}
in a generic moduli space of $\{t^a\}$. 
Here, $\mathbb{Z}_2^{\rm CP}$ transformation is given by Eqs. (\ref{eq:CPtrf}) and (\ref{eq:CPdef}).

\subsection{Small resolutions of $T^6/\mathbb{Z}_3$ orbifold}
\label{sec:smallblowZ3}

In this section, we incorporate the twisted (blow-up) modes in the effective action on the 6D toroidal orbifolds. 
We mainly deal with the blown-up $T^6/\mathbb{Z}_3$ orbifold, which is directly connected with the effective action of untwisted modes in the previous section in the blow-down limit, i.e., the vanishing blow-up radii. 
On the blown-up $T^6/\mathbb{Z}_3$ geometry, 
only the K\"ahler structure deformations remain on the blown-up $T^6/\mathbb{Z}_3$ background. 
The totally $h^{1,1}=36$ K\"ahler moduli include $h^{1,1}_{\rm untw}=9$ untwisted K\"ahler moduli and $h^{1,1}_{\rm tw}=27$ twisted K\"ahler moduli $c^d$ ($d=1,2,\cdots,27$) from the viewpoint of toroidal orbifold.\footnote{The following discussion is applicable to the T-dual of $T^6/\mathbb{Z}_3$ orbifold, where the massless deformations correspond to the complex structure moduli accompanying the $\widebar{27}_i$ matter fields. For more details, see, Ref. \cite{Ishiguro:2021csu} and references therein.} 
For simplicity, let us focus on the three diagonal untwisted K\"ahler moduli $t^a$ ($a=1,2,3$) whose moduli space is given by $\Pi_a SL(2,\mathbb{Z})_a$. 
The prepotential receives the corrections from the blow-up modes \cite{Dixon:1989fj}
\begin{align}
    G = t^1t^2t^3 -\frac{i}{4}\sum_{d=1}^{27}(c^d)^2,
    \label{eq:pre-3-2}
\end{align}
and the following vector
\begin{align}
    \begin{pmatrix}
       Y^A\\
       {\cal G}_A\\
    \end{pmatrix}
    =
    \begin{pmatrix}
    1\\
    t^1\\
    t^2\\
    t^3\\
    c^d\\
    2G-t^A\partial_AG\\
    t^2t^3\\
    t^1t^3\\
    t^1t^2\\
    -\frac{i}{2}c^d
    \end{pmatrix}
    =
    \begin{pmatrix}
    1\\
    t^1\\
    t^2\\
    t^3\\
    c^d\\
    -t^1 t^2 t^3\\
    t^2t^3\\
    t^1t^3\\
    t^1t^2\\
    -\frac{i}{2}c^d
    \end{pmatrix}    
,
\label{eq:vector_smallT6/Z3}
\end{align}
provides the K\"ahler potential
\begin{align}
\label{eq:Kahler-3-2}
    K_{\rm ks} =  -\ln \biggl[-i (\widebar{Y}^A{\cal G}_A -Y^A\widebar{{\cal G}}_A) \biggl] =
    -\ln \biggl[i(t^1-\widebar{t}^1)(t^2-\widebar{t}^2)(t^3-\widebar{t}^3) -\sum_d|c^d|^2 \biggl],
\end{align}
which is consistent with Ref. \cite{Ferrara:1989qb}. 
Correspondingly, the matter K\"ahler metric is calculated through Eq. (\ref{eq:KahlerMetric}). 

The allowed Yukawa couplings of matter zero-modes are restricted by $\mathbb{Z}_3$ symmetry. 
The matter superpotential takes the form
\begin{align}
W = A^1A^2 A^3 - \sum_{d,e,f=1}^{27}f(t^1,t^2,t^3)_{def} C^{d}C^{e}C^{f},
\end{align}
where $C^d$ denotes the twisted matters associated twisted K\"ahler moduli, and their 
Yukawa couplings $f(t^1,t^2,t^3)_{def}$ are moduli-dependent in general\cite{Hamidi:1986vh,Dixon:1986qv,Burwick:1990tu}. 
The holomorphic Yukawa couplings of untwisted modes enjoy the $S_4$ flavor symmetry as previously discussed in the orbifold regime, but the flavor symmetry of the twisted matter fields is complicated. 
Before going into the detail of the Yukawa couplings among twisted matters, 
we comment on the $S_4$ symmetry of twisted matter fields.
Let us denote them by fixed point labels, $C_{n_1,n_2,n_3}$, where $n_i=0,1,2$ represents the fixed point label 
on the $i$-th $T^2/\mathbb{Z}_3$.
The matter fields $C_{n,n,n}$ are obviously $S_4$ singlets.
The other $C_{n_1,n_2,n_3}$ are triplets because they transform each other under $S_4$.
Thus, $27$ twisted matter fields decompose into three $S_4$ singlets and eight $S_4$ triplets.

In the blow-down limit treated in this section, the relevant modular symmetry in the untwisted K\"ahler moduli space is $\Pi_{a=1}^3 SL(2,\mathbb{Z})_a \subset Sp(8,\mathbb{Z})$. 
It was known that the K\"ahler metrics of untwisted and twisted matters take the form \cite{Dixon:1989fj,Ibanez:1992hc}
\begin{align}
K_{a\bar{b}}^{(27)}\bigl|_{\rm untw} &= (K_{\rm ks})_{a\bar{b}}e^{-K_a} = \frac{1}{i(t^a -\bar{t}^a)} \delta_{a\bar{b}},
\nonumber\\
K_{d\bar{e}}^{(27)}\bigl|_{\rm tw} &= (K_{\rm ks})_{d\bar{e}}e^{-K_{\rm ks}/3}=\frac{1}{\left\{i(t^1-\bar{t}^1)(t^2-\bar{t}^2)(t^3-\bar{t}^3)\right\}^{2/3}} \delta_{d\bar{e}},
\end{align}
with $K_a = -\ln (-i (t^a -\bar{t}^a))$, from which we deduce the modular weight of both matter fields:
\begin{align}
A^1\,&:\, \left\{-1, 0, 0\right\},\quad A^2\,:\, \left\{ 0, -1, 0\right\},\quad A^3\,:\, \left\{0, 0, -1\right\},\quad C^d\,:\, \left\{-\frac{2}{3}, -\frac{2}{3}, -\frac{2}{3}\right\},   
\end{align}
under $\{SL(2,\mathbb{Z})_1, SL(2,\mathbb{Z})_2, SL(2,\mathbb{Z})_3\}$.
These modular weights play an important role in checking the modular invariance of the action. 
Indeed, the matter superpotential of untwisted modes $A^1A^2A^3$ is invariant under the 
$\Pi_a SL(2,\mathbb{Z})_a$ modular symmetry due to its modular weight -1 under all $SL(2,\mathbb{Z})_a$. 
However, 3-point couplings of twisted modes $C^dC^e C^f$ have modular weight $-2$ under all $SL(2,\mathbb{Z})_a$. Hence, the modular invariance requires the moduli-dependent couplings $f(t^1,t^2,t^3)_{def}$ with 
modular weight $1$ for each $SL(2,\mathbb{Z})_a$. 
It was known that the twisted modes belong to the triplet representation of double cover of $(A_4)_a$, namely $(T^\prime)_a \subset SL(2,\mathbb{Z})_a$ in the blow-down limit\cite{Lerche:1989cs,Ferrara:1989qb}, although this $T^\prime$ triplet is a reducible representation as seen below. 
Their holomorphic Yukawa couplings are expected to be a $\Gamma(3)$ modular form of non-Abelian discrete group $A_4$, which bring into the $(T^\prime)_a$-invariant superpotential. 
We will come back to this point later.

To see the $T^\prime$ structure in more detail, let us discuss $T^\prime$ transformations in the symplectic basis. 
Note that on this blown-up geometry, $S_4$ symmetry among untwisted modes lies in $Sp(8, \mathbb{Z}) \subset Sp(62, \mathbb{Z})$ in a way similar to the previous analysis. 
For concreteness, we focus on a subsector of twisted modes, i.e., three twisted fields $\{C^1,C^2, C^3\}$ which 
transform under the $SL(2,\mathbb{Z})_1$ symmetry. 
Then, the representations of these twisted matters under the $S_1$ and $T_1$ transformations of $SL(2,\mathbb{Z})_1$ 
are given by \cite{Ferrara:1989qb,Lerche:1989cs}
\begin{align}
\label{eq:S-R-tw}
    {\cal S}_{\rm tw} = -\frac{i}{\sqrt{3}}
    \begin{pmatrix}
    1 & 1 & 1 \\
    1 & w & w^2\\
    1 & w^2 & w 
    \end{pmatrix}
    ,
    \quad
    {\cal T}_{\rm tw} =
    \begin{pmatrix}
    w & 0 & 0\\
    0 & 1 & 0\\
    0 & 0 & 1
    \end{pmatrix}
    ,
\end{align}
with $w=e^{2\pi i/3}$. 
They satisfy the following algebraic relations:
\begin{align}
 {\cal S}_{\rm tw} ^2=-\mathbb{I},\qquad ( {\cal S}_{\rm tw}  {\cal T}_{\rm tw} )^3 =  {\cal T}_{\rm tw} ^3 = \mathbb{I} ,
\end{align}
which is the $T'$ symmetry  \cite{Ishimori:2010au,Ishimori:2012zz}.
Note that these matrices are rotated into
\begin{align}
    {\cal U}_{\rm tw}^{-1}{\cal S}_{\rm tw}{\cal U}_{\rm tw} = -\frac{i}{\sqrt{3}}
    \begin{pmatrix}
    1 & \sqrt{2} & 0 \\
    \sqrt{2} & -1 & 0\\
    0 & 0 & i 
    \end{pmatrix}
    ,
    \quad
     {\cal U}_{\rm tw}^{-1}{\cal T}_{\rm tw}{\cal U}_{\rm tw} = {\cal T}_{\rm tw},
\end{align}
by the unitary transformation:
\begin{align}
    {\cal U}_{\rm tw} =
    \begin{pmatrix}
    1 & 0 & 0\\
    0 & \frac{1}{\sqrt{2}} & -\frac{1}{\sqrt{2}}\\
    0 & \frac{1}{\sqrt{2}} & \frac{1}{\sqrt{2}}
    \end{pmatrix}
    .
\end{align}
It indicates that $T^\prime$ triplet of our interest is a reducible representation decomposed into 
the singlet and the doublet representations.

We find that these transformations lie in the symplectic modular group as follows:
 \begin{align}
 \begin{split}
S_{\rm tw}&=
\left(
    \begin{array}{cccccccccccc}
0 & -1 & 0 & 0 &           & 0 & 0 & 0 & 0 & \\
1 & 0 & 0 & 0  &           & 0 & 0 & 0 & 0 &\\
0 & 0 & 0 & 0 &            & 0 & 0 & 0 & -1&\\
0 & 0 & 0 & 0 &            & 0 & 0 & -1 & 0&\\
  &   &   &   & {\cal S}_{\rm tw} &   &   &    &  & \\
0 & 0 & 0 & 0 &            & 0 & -1 & 0 & 0 & \\
0 & 0 & 0 & 0 &            & 1 & 0 & 0 & 0 & \\
0 & 0 & 0 & 1 &            & 0 & 0 & 0 & 0 & \\
0 & 0 & 1 & 0 &            & 0 & 0 & 0 & 0 & \\
  &   &   &   &            &   &   &   &   & {\cal S}_{\rm tw}\\
\end{array}
  \right)
,
\\
T_{\rm tw}&=
\left(
    \begin{array}{cccccccccccc}
1 & 0 & 0 & 0 &            & 0 & 0 & 0 & 0 &\\
1 & 1 & 0 & 0 &            & 0 & 0 & 0 & 0 &\\
0 & 0 & 1 & 0 &            & 0 & 0 & 0 & 0 &\\
0 & 0 & 0 & 1 &            & 0 & 0 & 0 & 0 &\\
  &   &   &   & {\cal T}_{\rm tw} &   &   &   &   & \\
0 & 0 & 0 & 0 &            & 1 & -1 & 0 & 0&\\
0 & 0 & 0 & 0 &            & 0 & 1 & 0 & 0 &\\
0 & 0 & 0 & 1 &            & 0 & 0 & 1 & 0 &\\
0 & 0 & 1 & 0 &            & 0 & 0 & 0 & 1 &\\
  &   &   &   &            &   &   &   &   & {\cal T}_{\rm tw}\\
\end{array}
  \right)
,
 \end{split}
 \end{align}
where we omit 0 in the blank space for simplicity. 
Here, we focused on three twisted modes, but it is straightforward to extend this analysis on fully resolved toroidal orbifolds with small resolutions. 
In that case, the effective action enjoys the $(T^\prime)^3$ flavor symmetry belonging to $Sp(62,\mathbb{Z})$ modular symmetry.

Similar to the untwisted modes, $T^\prime$ flavor symmetry naturally lies in the symplectic modular flavor symmetry because it is a quotient of $SL(2,\mathbb{Z})$. 
The twisted modes have non-trivial representations under 
\begin{align}
 (T')^3\times S_4.   
\end{align}
Thus, their vacuum expectation values may lead non-trivial breaking of $(T')^3\times S_4$.
Together with the CP transformation as discussed in Sec. \ref{sec:unification}, symplectic modular symmetry is enlarged to the generalized symplectic modular symmetry. 
In the current setup, each $T^\prime$ flavor symmetry is enhanced to 
 \begin{align}
     T^\prime \rtimes \mathbb{Z}_2^{\rm CP} \simeq S_4,
 \end{align}
as an element of $GSp(62,\mathbb{Z})$. 
Note that $\mathbb{Z}_2^{\rm CP}$ transformation exchanges between the irreducible representations of $T^\prime$, i.e., the outer automorphism of $T'$.

To understand the nature of Yukawa couplings among twisted matters, 
we remind the couplings of twisted matters on $T^2/\mathbb{Z}_3$ 
geometry whose fixed points are labeled by $n=0,1,2$. 
The total 27 fixed points on $T^6/\mathbb{Z}_3$ orbifold is reproduced by the product of three fixed points 
on $T^2/\mathbb{Z}_3$. 
As a consequence of $\mathbb{Z}_3$ symmetry, Yukawa couplings of twisted modes localized on 
\begin{align}
(0,0,0),\quad (1,1,1),\quad (2,2,2),\quad (0,1,2)
\end{align}
are only allowed on $T^2/\mathbb{Z}_3$. 
In particular, their Yukawa couplings are of the form\cite{Hamidi:1986vh,Dixon:1986qv,Burwick:1990tu}
\begin{align}
f(t^1,t^2,t^3)_{def} \propto e^{-{\cal A}},
\end{align}
where ${\cal A}$ denotes the area satisfying boundary conditions relevant to twisted strings. 
Furthermore, these twisted strings wrap the cycle of toroidal orbifolds many times. 
Summing up these effects, the Yukawa couplings 
\begin{align}
f(t^1,t^2,t^3)_{def} \propto \sum_m e^{-{\cal A}_m}
\end{align}
will be elliptic functions such as theta functions as explicitly demonstrated on orbifolds. 
Note that the leading terms of holomorphic Yukawa couplings for $(n,n,n)$ with $n=0,1,2$ are 
constant, but the other one $(0,1,2)$ is exponentially suppressed with respect to the torus radii due to the 
fact that twisted strings propagate the fixed points with another. 
That is world-sheet instanton effects.
The couplings among the fixed points  $(0,1,2)$ would vanish if we neglect such world-sheet instanton effects.

Let us revisit these results in the context of symplectic modular symmetry, focusing on the modular transformation of Yukawa couplings on $T^2/\mathbb{Z}_3$.
We denote the Yukawa coupling among three twisted modes on the same fixed point by $Y_0$, while 
we denote the Yukawa couplings for three different fixed points (0,1,2) and (2,1,0) by $Y_1$ and $Y_2$, following 
Ref.~\cite{Lauer:1989ax}.
The triplet $(Y_0,Y_1,Y_2)$ transforms in the same way as the three twisted modes, i.e., Eq.~(\ref{eq:S-R-tw}), up to an overall phase.
Thus, they decompose the singlet and doublet under $T'$, but the singlet corresponds to $(Y_1-Y_2)$ and it vanishes automatically.
The doublet corresponds to the modular forms of weight 1 for $T'$.
Indeed, the modular forms of weight 1 for $T'$ have two degrees of freedom, corresponding to the $T'$ doublet \cite{Liu:2019khw}.
Appendix \ref{app} shows the modular form of weight 1 for $T'$ in a representation basis \footnote{
Similar to twisted modes and their Yukawa couplings, zero-modes and their Yukawa couplings in magnetized D-brane models have non-trivial behavior under the modular symmetry \cite{Kobayashi:2018rad,Kobayashi:2018bff,Ohki:2020bpo,Kikuchi:2020frp,Kikuchi:2020nxn,Kikuchi:2021ogn,Almumin:2021fbk,Tatsuta:2021deu}.}.
Indeed, this fact is well in accord with our observations. 
The constant Yukawa couplings localized at the fixed points $(n,n,n)$ with $n=0,1,2$ are the remnant of such modular functions in the large volume regime, whereas other one at $(0,1,2)$ are exponentially suppressed due to the structure of elliptic functions. The large volume expansion is nothing but the $q=e^{2\pi i t}$ expansion in the modular form. 
So far, we have focused on $T^2/\mathbb{Z}_3$ orbifold, but the structure discussed above holds for the Yukawa couplings of twisted modes associated with 27 fixed points on $T^6/\mathbb{Z}_3$ orbifold.

In the limit that the values of blow-up moduli are sufficiently large, $27$ twisted modes may have the flavor symmetry $S_{27}$ in $Sp(54,\mathbb{Z})$ modular symmetry as the model in Sec. \ref{sec:largeblowZ3Z3}.
However, in the blow-down orbifold limit, the symmetry $S_{27}$ as well as $Sp(54,\mathbb{Z})$ loses the geometrical meaning. 
However, twisted modes transform non-trivially under the $SL(2,\mathbb{Z})$ modular group of untwisted modes 
as the $T'$ reducible triplet because of the structure of prepotential (\ref{eq:pre-3-2}) and K\"ahler potential (\ref{eq:Kahler-3-2}).
That is a quite trivial transition of modular symmetric behavior of the twisted modes.
However, we have understood just two regimes, i.e., sufficiently large blow-up moduli regime and the blow-down orbifold limit. 
The middle regime is unclear in our analysis. 
We hope to report on the whole picture of modular symmetric behavior in the future.

Finally, we comment on the other untwisted moduli whose moduli space is given by $\frac{SU(3,3)}{SU(3)\times SU(3)\times U(1)}$. 
From the prepotential of the untwisted modes
\begin{align}
    G = \sum_{a,b,c=1}^9 \kappa_{abc}t^at^bt^c
\end{align}
with 
\begin{align}
    \kappa_{123}=\kappa_{468}=\kappa_{579}=1, \quad \kappa_{169}=\kappa_{258}=\kappa_{347}=-1,
\end{align}
and 0 otherwise, the K\"ahler potential of untwisted modes is provided by
\begin{align}
    K_{\rm ks}= - \ln \biggl[-i \frac{\kappa_{abc}}{6} (t^a - \widebar{t}^a)(t^b - \widebar{t}^b)(t^c - \widebar{t}^c)\biggl].
\end{align}
Here, we take the $Y^0 =1$ gauge. It is also known that 
the matter K\"aher potential and corresponding metric are given by\cite{Cvetic:1989ii}
\begin{align}
    K &= -\ln \det (-i({\cal T} - \widebar{{\cal T}})-A_\alpha A_{\bar{\alpha}})^{\hat{a}\bar{\hat{b}}} 
    = K_{\rm ks} + (-i({\cal T} - \widebar{{\cal T}}))^{-1}_{\hat{a}\bar{\hat{b}}}A^{\hat{a}}_\alpha A^{\bar{\hat{b}}}_{\bar{\alpha}} +{\cal O}(|A|^4),
    \nonumber\\
    K_{\hat{a}\bar{\hat{b}}}^{(27)} &= (-i({\cal T} - \widebar{{\cal T}}))^{-1}_{\hat{a}\bar{\hat{b}}},
\end{align}
where
\begin{align}
{\cal T}=
\begin{pmatrix}
t^1 & t^4 & t^5\\
t^7 & t^2 & t^6\\
t^8 & t^9 & t^3\\
\end{pmatrix}
,
\end{align}
with $\hat{a},\hat{b}=1,2,3$. 
Note that $\alpha$ stands for the indices of $SU(3)\subset E_8$ gauge group which are different from the indices of $SU(3)$ isometry $\hat{a}$. The superpotential of untwisted modes is written by\footnote{The kinetic mixing of matter fields will play an important role of realizing the hierarchical structure of matter fields \cite{Ishiguro:2021drk}.}
\begin{align}
W = \sum_{\hat{a},\hat{b},\hat{c},\alpha,\beta,\gamma}\kappa_{\hat{a}\hat{b}\hat{c}}\epsilon^{\alpha \beta \gamma}A^{\hat{a}}_\alpha A^{\hat{b}}_\beta A^{\hat{c}}_\gamma,
\end{align}
with $\kappa_{\hat{1}\hat{2}\hat{3}}=1$ and 0 otherwise, up to the overall factor, where $\epsilon^{\alpha \beta \gamma}$ are the anti-symmetric tensors of $SU(3)\subset E_8$ gauge group, and the upper and lower indices correspond to the $SU(3)$ isometry and $SU(3)\subset E_8$ gauge group, respectively. 
The $S_4$ flavor symmetry appears in three untwisted matters, e.g., $\{A_1^{\hat{1}}, A_1^{\hat{2}}, A_1^{\hat{3}}\}$ which can be embedded into the symplectic modular group following the previous procedures. 
In this way, the 9 untwisted modes transform as 3 triplet representations under the $S_4$ symmetry, that is, $\{ A_\alpha^{\hat{a}}, A_\alpha^{\hat{b}}, A_\alpha^{\hat{c}} \}$ for arbitrary $\alpha$.

\subsection{Small resolutions of $T^6/(\mathbb{Z}_3\times \mathbb{Z}_3^\prime)$ orbifold}
\label{sec:smallblowZ3Z3}

We deal with another example to exemplify the existence of generalized symplectic modular symmetry. 
We focus on the small resolutions of $T^6/(\mathbb{Z}_3\times \mathbb{Z}_3^\prime)$ orbifold by further imposing $\mathbb{Z}_3^\prime$ projection on $T^6/\mathbb{Z}_3$ orbifold. 
By introducing the complex coordinates of $\mathbb{C}^3$ as $z^m$, $m=1,2,3$, we identify them with 
\begin{align}
z^m \simeq z^m +1, \quad z^m \simeq z^m +\beta,
\end{align}
with $\beta =e^{2\pi i/6}$ to construct the three torus. 
Then, we specify the $\mathbb{Z}_3\times \mathbb{Z}_3^\prime$ action following Ref. \cite{Strominger:1985ku}:
\begin{align}
\mathbb{Z}_3\,&:\,Q(z^m) \simeq \beta^2 z^m,
\nonumber\\
\mathbb{Z}_3^\prime\,&:\,Q^\prime(z^m) \simeq \beta^{2m}z^m + \frac{1+\beta}{3},
\label{eq:orbproZ3Z3}
\end{align}
where $\mathbb{Z}_3^\prime$ is a freely acting symmetry. 
From 27 fixed points under $\mathbb{Z}_3$ symmetry
\begin{align}
    z_{\rm fix} = \frac{1+\beta}{3}(n_1, n_2, n_3),
\end{align}
with $n_m=0,1,2$, we find 9 orbifold-invariant combinations: 
\footnotesize
\begin{align}
    &f_1:(0,0,0)+(1,1,1)+(2,2,2),\,\, f_2:(0,1,2)+(1,2,0)+(2,0,1),\,\, f_3:(0,2,1)+(1,0,2)+(2,1,0), 
    \label{eq:fixed1}
    \\
    &f_4:(1,0,0)+(2,1,1)+(0,2,2),\,\, f_5:(0,1,0)+(1,2,1)+(2,0,2),\,\, f_6:(0,0,1)+(1,1,2)+(2,2,0), 
    \label{eq:fixed2}
    \\
    &f_7:(2,0,0)+(0,1,1)+(1,2,2),\,\, f_8:(0,2,0)+(1,0,1)+(2,1,2),\,\, f_9:(0,0,2)+(1,1,0)+(2,2,1).
    \label{eq:fixed3}
\end{align}
\normalsize
Hence, there exist $h^{1,1}_{\rm tw}=9$ twisted K\"ahler moduli in addition to $h^{1,1}_{\rm untw}=3$ untwisted K\"ahler moduli. 

The orbifold projection (\ref{eq:orbproZ3Z3}) breaks $(T^\prime)^3$ symmetry in Sec. \ref{sec:smallblowZ3} to $S_3$. 
For instance, Eq. (\ref{eq:fixed1}) represents the $S_3$ singlet, and Eqs. (\ref{eq:fixed2}) and (\ref{eq:fixed3}) become $S_3$ triplets, although the triplet representation is the reducible representation decomposed into the single and doublet.
In addition, $27$ twisted modes on $T^6/\mathbb{Z}_3$ correspond to three singlets and eight triplets under the $S_4$ flavor symmetry of untwisted modes, where the matter fields $C_{n,n,n}$ correspond to the $S_4$ singlets, while the other $C_{n_1,n_2,n_3}$ are $S_4$ triplets.
All of the matter fields $C_{n,n,n}$  correspond to three fixed points in the linear combination $f_1$. 
Here, the matter fields corresponding to both $f_2$ and $f_3$ become $S_4$ singlets because of the orbifold projection (\ref{eq:orbproZ3Z3}) .
On the other hand, the matter fields corresponding to $(f_4,f_5,f_6)$ and $(f_7,f_8,f_9)$ still represent 
the $S_4$ triplets.
Thus, the orbifold projection (\ref{eq:orbproZ3Z3})  reduces three $S_4$ singlets and eight $S_4$ triplets to 
three $S_4$ singlets and two $S_4$ triplets. 
From the total $Sp(2\times 12+2, \mathbb{Z})$ symplectic group, 
we can derive the $S_3$ flavor group,
\begin{align}
    Sp(26, \mathbb{Z}) \supset G_{\rm flavor}=S_3,
\end{align}
where the embedding $S_3\subset Sp(8,\mathbb{Z})\subset Sp(26,\mathbb{Z})$ is realized by following the same step in the previous section. The $S_3$ generator corresponds to $P$ in Eq. (\ref{eq:PQ}). 
Then, the $S_3$ flavor symmetry is enlarged to $D_6\simeq S_3 \rtimes \mathbb{Z}_2^{\rm CP}\subset GSp(26,\mathbb{Z})$, taking into account the CP symmetry. 
Here, $\mathbb{Z}_2^{\rm CP}$ transformation is given by Eqs. (\ref{eq:CPtrf}) and (\ref{eq:CPdef}).

In the following, we move to the geometrical regime where the size of blow-up radii is above the string length and check whether there exist such discrete modular flavor symmetries.

\subsection{Geometrical regime of blown-up $T^6/(\mathbb{Z}_3\times \mathbb{Z}_3^\prime)$ orbifold}
\label{sec:largeblowZ3Z3}
In the previous sections, we examined the blown-up toroidal orbifolds with small resolutions, where the $S_3$, $S_4$, and $T^\prime$ flavor symmetries among untwisted and twisted matters are independent. 
Together with the CP symmetry, all of them belong to the generalized symplectic modular symmetry and are enlarged to $D_6$, $S_4\rtimes \mathbb{Z}_2^{\rm CP}$ and $S_4$, respectively. 

In contrast to the previous analysis assuming the small vacuum expectation values of blow-up modes, we discuss more non-trivial examples on smooth CY threefolds. In this section, we focus on $T^6/(\mathbb{Z}_3\times \mathbb{Z}_3^\prime)$ background, where the blown-up cycles are larger than 1 in string units, in comparison with the analysis of Sec. \ref{sec:smallblowZ3Z3}. 
Among $h^{1,1}=12$ K\"ahler moduli, there exist $h^{1,1}_{\rm untw}=3$ untwisted K\"ahler moduli spanned by $e_a$ corresponding to $idz^a\wedge d\widebar{z}^{\widebar{b}}$ from the viewpoint of toroidal orbifold, where $z^a$ ($a=1,2,3$) denote the coordinates of 3-torus. 
In addition, we have $h^{1,1}_{\rm tw}=9$ twisted K\"ahler moduli spanned by $e_r$ $(r=1,2,...,9)$, corresponding to complex projective spaces $\mathbb{P}^2$'s, where fixed points are now resolved as $\mathbb{P}^2$ of a $\mathbb{C}^3/\mathbb{Z}_3$ singularity. 
On this geometry, there is no complex structure deformation.

When we expand the K\"ahler form
\begin{align}
    J = \sum_{a=1}^3t^a e_a+\sum_{r=1}^9v^r e_r,
\end{align}
the K\"ahler potential is evaluated as \cite{DeWolfe:2005uu}
\begin{align}
\begin{split}
        K_{\rm ks} &= -\ln \int J_c \wedge J_c \wedge J_c = -\ln \frac{1}{8}\biggl[ 81(t^1-\widebar{t}^1)(t^2-\widebar{t}^2)(t^3-\widebar{t}^3)-\frac{3}{2}\sum_r (v^r-\widebar{v}^r)^3 \biggl].
        \label{eq:K_largeZ3Z3}
\end{split}
\end{align}
Here, the prepotential is given by
\begin{align}
    G = \frac{\kappa_{ABC}}{6} t^At^Bt^C = 81t^1 t^2 t^3 -\frac{3}{2}\sum_r(v^r)^3,
\end{align}
with the nonvanishing triple intersection numbers,
\begin{align}
    \kappa_{123}=81,\quad
    \kappa_{rrr}=-9.
\end{align}
Hence, the corresponding matter superpotential takes the form
\begin{align}
    W = 81 A^1A^2A^3 -9\sum_{r=1}^9 (V^r)^3,
\end{align}
where $A^{1,2,3}$ and $V^r$ denote the matter fields 
associated with the untwisted and twisted moduli fields from the viewpoint of toroidal orbifolds, respectively. 
The matter K\"ahler metric is calculated through Eq. (\ref{eq:KahlerMetric}). 
We find that the matter superpotential is invariant under two 
flavor symmetries: $S_4$ symmetry for the untwisted matters $\{A^1, A^2, A^3\}$ and $S_9$ symmetry for the twisted matters $\{V^1,V^2,...,V^{9}\}$ in addition to the overall modular symmetry $SL(2,\mathbb{Z})_{\rm overall}$.

To see the relation between flavor symmetries of matter fields and symplectic modular symmetry, 
let us consider the period vector providing the K\"ahler potential (\ref{eq:K_largeZ3Z3}). 
Since the blown-up $T^6/(\mathbb{Z}_3\times \mathbb{Z}_3^\prime)$ manifold have totally 12 K\"ahler moduli fields, the moduli space is described by $Sp(26,\mathbb{Z})$ symplectic modular symmetry. 
Correspondingly, the period vector is given by
\begin{align}
    \begin{pmatrix}
       Y^A\\
       {\cal G}_A\\
    \end{pmatrix}
    =
    \begin{pmatrix}
    1\\
    t^1\\
    t^2\\
    t^3\\
    v^r\\
    2G-t^A\partial_AG\\
    81t^2t^3\\
    81t^1t^3\\
    81t^1t^2\\
    -\frac{9}{2}(v^r)^2
    \end{pmatrix}
    =
    \begin{pmatrix}
    1\\
    t^1\\
    t^2\\
    t^3\\
    v^a\\
    -81t^1 t^2 t^3+\frac{3}{2}\sum_r (v^r)^3\\
    81t^2t^3\\
    81t^1t^3\\
    81t^1t^2\\
    -\frac{9}{2}(v^r)^2
    \end{pmatrix}    
.
\end{align}

Similar to the previous example, the $S_4$ flavor 
symmetry of the untwisted moduli $\{t^1,t^2,t^3\}$ can be embedded into $Sp(8,\mathbb{Z}) \subset Sp(26, \mathbb{Z})$. 
Accordingly, the matter fields $A^{1,2,3}$ transform as in Eq. (\ref{eq:A4Trf}). 
In addition, we find that the generators of $S_9$ 
\begin{align}
    {\cal P}_{rs} : v^r \leftrightarrow v^s
\end{align}
is the element of $Sp(26, \mathbb{Z})$ as follows. 
For an illustrative purpose, let us focus on ${\cal P}_{12}$ exchanging 
$v^1$ and $v^2$. 
To shorten the expression of the generator ${\cal P}_{12}$, 
we focus on the relevant period vector transformed under ${\cal P}_{12}$,
\begin{align}
        \begin{pmatrix}
    v^1\\
    v^2\\
    -\frac{9}{2}(v^1)^2\\
    -\frac{9}{2}(v^2)^2
    \end{pmatrix}   
    ,
\end{align}
on which ${\cal P}_{12}$ acts on 
\begin{align}
\begin{split}
\begin{pmatrix}
0 & 1 & 0 & 0 \\
1 & 0 & 0 & 0 \\
0 & 0 & 0 & 1 \\
0 & 0 & 1 & 0 \\
\end{pmatrix}
        \begin{pmatrix}
    v^1\\
    v^2\\
    -\frac{9}{2}(v^1)^2\\
    -\frac{9}{2}(v^2)^2
    \end{pmatrix}   
    =
            \begin{pmatrix}
    v^2\\
    v^1\\
    -\frac{9}{2}(v^2)^2\\
    -\frac{9}{2}(v^1)^2
    \end{pmatrix}   
    .
 \end{split}
\label{eq:P12}
 \end{align}
${\cal P}_{12}$ acts the identity operators on the other elements of period vector. 
In this way, we show that the generator ${\cal P}_{12}$ is 
the element of $Sp(26,\mathbb{Z})$, and it is possible to prove the other ${\cal P}_{rs}$ as the element of $Sp(26,\mathbb{Z})$ by changing the indices of $v^{r,s}$. 

These transformations are directly related to the $S_9$ flavor symmetry of matter fields from ${\cal P}_{rs}$. 
Indeed, the submatrix of ${\cal P}_{rs}$ acting on $(v^1,v^2)^T$ 
\begin{align}
\begin{pmatrix}
0 & 1 \\
1 & 0 \\
\end{pmatrix}
\label{eq:S9Trf}
 \end{align}
corresponds to the transformation between $V^1$ and $V^2$. 
Combining all ${\cal P}_{ab}$ acting on 
\begin{align}
(v^1,v^2,v^3,v^4,v^5,v^6,v^7,v^8,v^9)^T,    
\end{align}
one can realize the transformations of matter fields under $S_9$ modular flavor symmetry. 

We conclude that there exist $G_{\rm flavor}=S_4\times S_9$ flavor symmetries in the effective action of blown-up $T^6/(\mathbb{Z}_3\times \mathbb{Z}_3^\prime)$ orbifold. 
They originate from the $Sp(26, \mathbb{Z})$ symplectic group, 
\begin{align}
    Sp(26, \mathbb{Z}) \supset G_{\rm flavor}=S_4\times S_9,
\end{align}
where $S_9$ symmetry is broken down to $S_3$ in the regime with small resolutions as in Sec. \ref{sec:smallblowZ3Z3}. 
That would be related to the spontaneous breaking of flavor symmetry triggered by the blow-up modes. 
We will leave the more detailed discussion of this symmetry breaking involving the stabilization of blow-up modes to future work. 
When we take into account the CP transformation, 
two flavor symmetries are enhanced to 
\begin{align}
    S_4 \rtimes \mathbb{Z}_2^{\rm CP},
    \quad
    S_9 \rtimes \mathbb{Z}_2^{\rm CP}.
\end{align}
In the next section, we show the existence of non-Abelian discrete flavor symmetries on other classes of CY threefolds.

This example suggests that some class of models with $h$ blow-up modes may lead to the 
superpotential
\begin{align}
W=\sum_{i=1}^h (A^i)^3,
\end{align}
up to an overall factor.
This superpotential seems to have the $S_h$ flavor symmetry, which is 
the permutation among $A^i$.
This type of superpotential has a specific property for $h=3$.
We start with the following superpotential:
\begin{align}
W=\tilde A^1 \tilde A^2 \tilde A^3.
\end{align}
Here, we redefine the matter fields,
\begin{align}
\tilde A^1&=A^1+A^2+A^3, \notag \\
\tilde A^1&=A^1+\omega A^2+\omega^2 A^3, \\
\tilde A^1&=A^1+\omega^2 A^2+ \omega A^3.\notag 
\end{align}
Then, we can write 
\begin{align}
W=\tilde A^1 \tilde A^2 \tilde A^3 = (A^1)^3+(A^2)^2+(A^3)^2.
\end{align}
The $S_3$ symmetry is enlarged to $S_4$. 
Hence, this phenomenon can be observed only for $h =3$.

\subsection{Three-parameter examples of Calabi-Yau threefolds}
\label{sec:CY}

So far, we have focused on the toroidal orbifolds with and without resolutions. 
In this section, we analyze the smooth CY threefolds, emphasizing the flavor symmetry of matter fields. 
In particular, we deal with the Kreuzer-Skarke data set of CY threefolds, given by total 473,800,776 ambient toric fourfolds \cite{Kreuzer:2000xy}. 
The ambient toric varieties are described by the corresponding (reflexive) polytope, which might involve 
singularities in general. The smooth CY hypersurfaces are obtained through ``triangulation'' of the polytope where 
the singularities of the ambient space are resolved. 
For the favorable CY hypersurface $X$\footnote{The favorable CY means that the hodge number $h^{1,1}$ of $X$ is the same with that of the ambient space.}, the harmonic (1,1)-forms for $H^2(X, \mathbb{Z})$, $J_r$ with $r=1,2,\cdots,h^{1,1}$, can be obtained by the pull-backs of (1,1)-forms on the ambient space. Then, the intersection numbers $d_{rst}$ of $X$ can be calculated by
\begin{align}
d_{rst} = \int_X J_r \wedge J_s \wedge J_t.
\end{align}
Here, the $J_r$ stand for the basis of the Picard group. 
Finally, we describe the K\"ahler cone of $X$, determining the allowed K\"ahler moduli $t^r$. 
For the favorable CY hypersurface, the K\"ahler cone of $X$ is also derived from that of the 
ambient space whose K\"ahler cone is calculated by the data of polytope. 
When we expand the (1,1)-form $J$ as $J=t^r J_r$, the K\"ahler moduli $t^r$ satisfy
\begin{align}
\sum_r K^{\hat{r}}_{r}t^r \geq 0,
\label{eq:Kcone}
\end{align}
where we introduce the K\"ahler cone matrix $K=[K^{\hat{r}}_{r}]$ and $\hat{r}$ runs over the facets of the K\"ahler cone. 
Note that the two examples we analyze later have a few different triangulation corresponding to a partial resolution for the ambient toric varieties. In that case, the K\"ahler cone is given by the union of several sub-cones:
\begin{align}
    K = \bigcup_{j} K_j.
\end{align}
 
Among them, only 16 manifolds have a non-trivial fundamental group which is an important ingredient in introducing Wilson-lines on quotient CY threefolds \cite{Batyrev:2005jc}. 
We focus on these particular class of CY threefolds in the context of heterotic 
string theory with standard embedding.\footnote{The heterotic string models in the context of non-standard embedding are discussed on 
such a particular class of CY threefolds \cite{He:2013ofa}.}
For illustrative purposes, we deal with five quotient CY threefolds. (For more details about the toric data as well as topological data of CY threefolds, we refer to Ref. \cite{He:2013ofa}.) In particular, we focus on the K\"ahler moduli sector by assuming that the complex structure moduli sector is stabilized by some mechanism at a high scale. 

\begin{enumerate}
\item 
The first example of CY threefold with non-trivial fundamental group 
has $h^{1,1}=3$ K\"ahler moduli,  $h^{2,1}=59$ complex 
structure moduli, i.e., Euler number $\chi =-112$ (corresponding to the ``downstairs'' CY threefold $X_{14}$ 
in Ref. \cite{He:2013ofa}). 
We will focus on the K\"ahler moduli sector since the complex structure moduli sector is complicated to analyze. 
The prepotential in the K\"ahler moduli sector is given by
\begin{align}
G = t^1t^2t^3, 
\end{align}
with non-vanishing triple intersection number $\kappa_{123}=1$. 
The K\"ahler cone is given by $K={\rm diag}(1,1,1)$. 
Hence, the superpotential of matter fields associated with the K\"ahler moduli 
has the same functional form as one in the example of Sec. \ref{sec:smallblowZ3},
\begin{align}
W = A^1A^2A^3.
\end{align}
Hence, in addition to $\Pi_{a=1}^3SL(2,\mathbb{Z})_a$, 
there exists the $S_4$ flavor symmetry, which can be enhanced to $S_4\rtimes \mathbb{Z}_2^{CP}$ 
symmetry, taking into account the CP transformation. 
The symmetry is then embedded into the generalized symplectic $Sp(8,\mathbb{Z})$ flavor symmetry in the K\"ahler moduli space. 

\item 
The second example of three K\"ahler parameters of CY has $h^{2,1}=59$ complex 
structure moduli, i.e., Euler number $\chi =-112$ (corresponding to the downstairs CY threefold $X_{6}$ 
in Ref. \cite{He:2013ofa}). 
When we focus on the K\"ahler moduli sector, the prepotential is given by
\begin{align}
G = t^1t^2t^3 - 2 t^2(t^3)^2, 
\end{align}
with non-vanishing triple intersection numbers $\kappa_{123}=1$ and $\kappa_{233}=-4$. 
Here, the intersection number is negative, but it is possible to redefine the 
modulus field as $\widetilde{t}^1\equiv t^1 -2t^3$, corresponding to the K\"ahler cone $K={\rm diag}(1,1,1)$. 
It provides the superpotential of matter fields associated with the K\"ahler moduli,
\begin{align}
W = A^1A^2A^3 - 2 A^2(A^3)^2,
\end{align}
which has a flavor structure similar to the first example of CY. 
Indeed, when we redefine the matter fields as $\widetilde{A}^1\equiv A^1 -2A^3$, 
the matter superpotential reduces to
\begin{align}
W = \widetilde{A}^1A^2 A^3.    
\end{align}
Hence, we find that $S_4$ flavor symmetry is realized on this class of CY threefold and its enhanced $S_4\rtimes \mathbb{Z}_2^{CP}$ symmetry together with the CP transformation can be embedded into the generalized symplectic $Sp(8,\mathbb{Z})$ flavor symmetry. 

\item
The third example of three K\"ahler parameters of CY has $h^{2,1}=43$ complex 
structure moduli, i.e., Euler number $\chi =-80$ (corresponding to the downstairs CY threefold $X_{5}$ 
in Ref. \cite{He:2013ofa}). 
The prepotential in the K\"ahler moduli sector is given by
\begin{align}
G = t^1t^2t^3 - 2 t^1(t^3)^2 +2 t^2(t^3)^2 - 8 (t^3)^3, 
\end{align}
with non-vanishing triple intersection numbers $\kappa_{123}=1$, $\kappa_{133}=-4$, $\kappa_{233}=4$ 
and $\kappa_{333}=-48$. 
Here, the K\"ahler cone is given by
\begin{align}
    K = 
    \begin{pmatrix}
    0 & 1 & -2\\
    1 & 0 & 0 \\
    0 & 0 & 1\\
    \end{pmatrix}
    ,
\end{align}
constraining the K\"ahler moduli to satisfy Eq. (\ref{eq:Kcone}). 
It provides the superpotential of matter fields associated with the K\"ahler moduli,
\begin{align}
W = A^1A^2A^3 - 2 A^1(A^3)^2 +2 A^2(A^3)^2 - 8 (A^3)^3.
\end{align}
There is the overall modular symmetry $SL(2,\mathbb{Z})_{\rm overall}$.
It seems that other non-trivial flavor symmetries are absent in this system.  
There is $\mathbb{Z}_2$ symmetry such as 
\begin{align}
A^1 \to A^2, \qquad A^2 \to A^1, \qquad A^3 \to -A^3,
\end{align}
up to $U(1)_R$ symmetry $W \to -W$.

\item
The fourth example of three K\"ahler parameters of CY has $h^{2,1}=75$ complex 
structure moduli, i.e., Euler number $\chi =-144$ (corresponding to the downstairs CY threefold $X_{7}$ 
in Ref. \cite{He:2013ofa}). 
The prepotential in the K\"ahler moduli sector is given by
\begin{align}
G = t^1t^2t^3 - 2 (t^2)^2t^3  - 2 t^1(t^3)^2 + 8 (t^3)^2, 
\end{align}
with non-vanishing triple intersection numbers $\kappa_{123}=1$, $\kappa_{223}=\kappa_{133}=-4$ 
and $\kappa_{333}=48$. 
Here, the K\"ahler cone is given by
\begin{align}
    K = 
    \begin{pmatrix}
    1 & -2 & 0\\
    0 & 0 & 1 \\
    0 & 1 & -2\\
    \end{pmatrix}
    ,
\end{align}
constraining the K\"ahler moduli to satisfy Eq. (\ref{eq:Kcone}).
It provides the superpotential of matter fields associated with the K\"ahler moduli,
\begin{align}
W = A^1A^2A^3 - 2 (A^2)^2A^3 - 2 A^1(A^3)^2 + 8 (A^3)^2.
\end{align}
There is the overall modular symmetry $SL(2,\mathbb{Z})_{\rm overall}$, 
but other non-trivial flavor symmetries are absent in this system.

\item
The last example of three K\"ahler parameters of CY has $h^{2,1}=75$ complex 
structure moduli, i.e., Euler number $\chi =-144$ (corresponding to the downstairs CY threefold $X_{8}$ 
in Ref. \cite{He:2013ofa}). 
The prepotential in the K\"ahler moduli sector is given by
\begin{align}
G = t^1t^2t^3 - 2 t^1(t^3)^2 -2 t^2(t^3)^2 + 8 (t^3)^2, 
\end{align}
with non-vanishing triple intersection numbers $\kappa_{123}=1$, $\kappa_{133}=\kappa_{233}=-4$ 
and $\kappa_{333}=48$. 
Here, the K\"ahler cone is given by
\begin{align}
    K = 
    \begin{pmatrix}
    0 & 0 & 1\\
    1 & 0 & -2 \\
    0 & 1 & -2\\
    \end{pmatrix}
    ,
\end{align}
constraining the K\"ahler moduli to satisfy Eq. (\ref{eq:Kcone}). 
It provides the superpotential of matter fields associated with the K\"ahler moduli,
\begin{align}
W = A^1A^2A^3 - 2 A^1(A^3)^2 -2 A^2(A^3)^2 + 8 (A^3)^2. 
\end{align}
There is the overall modular symmetry $SL(2,\mathbb{Z})_{\rm overall}$.
In addition, there exists a $\mathbb{Z}_2\simeq S_2$ flavor symmetry between $A^1$ and $A^2$ and 
it is further enhanced to $\mathbb{Z}_2\rtimes \mathbb{Z}_2^{CP}$ 
symmetry, taking into account the CP transformation. 
Since the $\mathbb{Z}_2$ transformation is the same with ${\cal P}_{12}$ in Eq. (\ref{eq:P12}), 
the symmetry is then embedded into the $Sp(8,\mathbb{Z})$ generalized symplectic flavor symmetry. 

\end{enumerate}

We find that non-trivial flavor symmetries on three-parameter examples of CY threefolds are given by 
$S_4$ and $\mathbb{Z}_2$ symmetries, enlarging to $S_4\rtimes \mathbb{Z}_2^{\rm CP}$ and $\mathbb{Z}_2 \rtimes \mathbb{Z}_2^{\rm CP}$ taking into account the 
$\mathbb{Z}_2^{\rm CP}$ symmetry. 
Hence, the $S_4$ flavor symmetry appears in not only the exact orbifold limit but also a certain class of CY threefolds such as the first 
and second examples.
In addition, all examples enjoy the overall modular symmetry $SL(2,\mathbb{Z})_{\rm overall}$, which 
looks $\mathbb{Z}_3$ symmetry, $A^i \to e^{2\pi i/3}A^i$ in the field-theoretical viewpoint. 
The above $S_4$ and $\mathbb{Z}_2$ symmetries are commutable with this $\mathbb{Z}_3$ symmetry.

Our analysis is limited to specific CY threefolds, but it is interesting to examine the flavor symmetries on a more broad class of CY threefolds. We leave it for future work. 

\section{Conclusions and discussions}
\label{sec:con}

In this paper, we examined the origin of flavor symmetry in the context of heterotic string theory on CY threefolds, including toroidal orbifolds. 
The flavor structure of matter zero-modes in the context of heterotic string theory with standard embedding is 
governed by the geometrical structure of CY threefolds because the matter and moduli fields are in one-to-one correspondence with each other. 
It indicates that the flavor structure is closely related to the symplectic structure of CY moduli spaces. 
Furthermore, the 4D CP and the $U(1)_R$ symmetries were regarded as the outer automorphism of the symplectic modular group \cite{Ishiguro:2020nuf} and the rotation of holomorphic three-form of CY threefolds \cite{Witten:1985xc}, respectively. 
These observations motivate us to thoroughly investigate the relation among flavor, CP, and $U(1)_R$ symmetries on the basis of 4D effective action. 
This approach would yield the ultra-violet origin of models with modular flavor symmetry in the bottom-up approach. 

In Sec. \ref{sec:2}, we gave a unified picture of these flavor, CP, and $U(1)_R$ symmetries from the viewpoint of  
generalized symplectic symmetry $GSp(2h+2, \mathbb{C})\simeq Sp(2h+2,\mathbb{C})\rtimes \mathbb{Z}_2^{\rm CP} \times U(1)_R$, with $h$ being the number of either complex structure moduli or K\"ahler moduli. 
It was known that the traditional flavor symmetry was reflected by the flavor structure of Yukawa couplings, 
but our results suggest that the flavor symmetry $G_{\rm flavor}$ is enlarged to some non-Abelian groups $G_{\rm flavor}\rtimes \mathbb{Z}_2^{\rm CP}$ by the CP symmetry on a general class of CY threefolds with standard embedding. 
To exemplify these results, we deal with toroidal orbifolds with and without resolutions, and three-parameter examples of CY threefolds. 
It turned out that three untwisted modes associated with three K\"ahler moduli enjoy $S_4$ flavor symmetry on toroidal orbifolds. 
On the $T^6/\mathbb{Z}_3$ and $T^6/(\mathbb{Z}_3\times \mathbb{Z}_3^\prime)$ orbifolds with small resolutions, 
there exist $(T^{\prime})^3$ and $S_3$ symmetries among twisted modes, respectively. 
On the geometrical regime of $T^6/(\mathbb{Z}_3\times \mathbb{Z}_3^\prime)$ orbifold where the size of blow-up modes is larger than the string length, we find the $S_9$ flavor symmetry among blow-up models (corresponding to twisted modes in the blow-down limit). 
The three-parameter examples of CY threefolds also enjoy $\mathbb{Z}_2$ and $S_4$ flavor symmetries as discussed in Sec. \ref{sec:CY}. 
Hence, non-trivial flavor symmetries such as  $S_4$  appear in not only the exact orbifold limit but also a certain class of CY threefolds.
All the flavor symmetries we analyzed have an origin in the symplectic modular symmetry and are enlarged to non-Abelian groups by the CP symmetry.

In this paper, we have proposed the unification of flavor, CP, and $U(1)_R$ symmetries originated from the symplectic modular 
symmetry, and exemplify the flavor symmetries on CY threefolds, including toroidal results. 
However, there exist several future directions to pursue. 
\begin{itemize}

\item Yukawa couplings on CY threefolds

On the toroidal backgrounds, moduli-dependent Yukawa couplings are estimated by the modular weight of matter fields, 
and their functional form is described by a certain modular form of $SL(2,\mathbb{Z})$ symmetry. 
On the other hand, the results of Sec. \ref{sec:CY} exhibit the existence of $S_4$ flavor 
symmetry of matter fields on some CY threefolds equipping the effective $SL(2,\mathbb{Z})$ modular symmetry. 
Hence, the Yukawa couplings including quantum corrections would be described by the modular form of $S_4$ 
flavor symmetry, controlled by the underlying symplectic modular symmetry. 
In this paper, we have focused on the CY moduli spaces at the classical level, 
but loop and instanton corrections calculated in the topological string theory appear in the prepotential as well as Yukawa couplings \cite{Gopakumar:1998ii}. Also, massive modes generate the higher-order couplings, which are described in an appropriate derivative of the prepotential \cite{Bershadsky:1993cx}. It would be fascinating to figure out the flavor structure of matter fields, including these corrections. 
We hope to report on this interesting direction in the future.

\item Spontaneous breaking of flavor symmetry

From the results in Secs. \ref{sec:smallblowZ3Z3} and \ref{sec:largeblowZ3Z3}, there the $S_3$ and $S_9$ flavor symmetries of twisted modes on $T^6/(\mathbb{Z}_3\times \mathbb{Z}_3^\prime)$ orbifold 
with small resolutions and its geometrical regime, respectively. 
It indicates the spontaneous breaking of flavor symmetry would be induced by the blow-up modes. To understand the underlying structure, it is required to interpolate both the orbifold and geometrical regime. 
Including the stabilization of blow-up modes, 
we leave this topic for future research.

\item Modular symmetric behavior of the twisted modes

In the limit that the size of blow-up moduli is sufficiently large, 
9 twisted modes in the example of Sec. \ref{sec:largeblowZ3Z3} have $S_9$ flavor symmetry in 
$Sp(18,\mathbb{Z})$ modular symmetry.
Similarly, 27 twisted modes in the example of Sec. \ref{sec:smallblowZ3} 
may also have the $S_{27}$ flavor symmetry in $Sp(54,\mathbb{Z})$ modular symmetry.
On the other hand, the symmetry $S_{27}$ as well as $Sp(54,\mathbb{Z})$ loses the geometrically meaning in the blow-down orbifold limit. 
However, twisted modes transform non-trivially under the $SL(2,\mathbb{Z})$ modular symmetry of untwisted modes 
because of the structure of prepotential and K\"ahler potential.
That is a quite trivial transition of modular symmetric behavior of the twisted modes. 
In this paper, we have understood just two regimes, i.e., sufficiently large blow-up moduli regime $({\rm Im}(v^r) \gg 1)$ and the blow-down orbifold limit $(|c^r| \ll 1)$ in the context of symplectic modular symmetry. 
The middle regime is unclear in our analysis. 
It was conjectured by Ref. \cite{GrootNibbelink:2009wzz} 
that the blow-up moduli in both regimes would be related by 
$c^r\simeq \exp (2\pi i v^r)$. It is interesting to find out a similar relation of twisted matters, which brings us to understand the whole picture of modular symmetric behavior. 

\end{itemize}

\subsection*{Acknowledgements}

The authors would like to thank Hiroshi Ohki for useful discussions, and especially to the anonymous referee for a careful reading of the manuscript and constructive suggestions.
H. O. was supported in part by JSPS KAKENHI Grant Numbers JP19J00664 and JP20K14477.

\appendix

\section{Modular form of weight 1 for $T'$}
\label{app}

Following Ref. \cite{Liu:2019khw}, we define the following functions,
\begin{eqnarray}
\hat e_1(\tau)=\frac{\eta^3(3\tau)}{\eta(\tau)},\qquad \hat e_2(\tau)=\frac{\eta^3(\tau/3)}{\eta(\tau)},
\end{eqnarray}
by using the Dedekind eta function $\eta(\tau)$, 
\begin{eqnarray}
\eta(\tau) = q^{1/24}\prod^\infty_{n=1}(1-q^n),
\end{eqnarray}
with $q=e^{2\pi i \tau}$.
Under the $T$-transformation, they transform 
\begin{eqnarray}
\hat e_1 \rightarrow \omega \hat e_1, \qquad 
\hat e_2 \rightarrow 3(1-\omega) \hat e_1 + \hat e_2,
\end{eqnarray}
with $w=e^{2\pi i/3}$. Under the $S$-transformation, 
they transform
\begin{eqnarray}
\hat e_1 \rightarrow 3^{-3/2}i (-\tau)\hat e_2, \qquad 
\hat e_2 \rightarrow  3^{3/2}i (-\tau)\hat e_1.
\end{eqnarray}

By using these functions, we can construct 
the doublet of $T'$, which transforms under the $T$-transformation,
\begin{eqnarray}
\left(\begin{array}{c}
x_1 \\ y_1
\end{array}
\right)
\rightarrow 
\rho(T) \left(\begin{array}{c}
x_1 \\ y_1
\end{array}
\right),
\end{eqnarray}
and under the $S$-transformation
\begin{eqnarray}
\left(\begin{array}{c}
x_1 \\ y_1
\end{array}
\right)
\rightarrow (-\tau)
\rho(S) \left(\begin{array}{c}
x_1 \\ y_1
\end{array}
\right),
\end{eqnarray}
where 
\begin{eqnarray}
\rho(T)=\left(
\begin{array}{cc}
\omega & 0 \\
0 & 1
\end{array}
\right), \qquad
\rho(S)=-\frac{i}{\sqrt 3}\left(
\begin{array}{cc}
1 & \sqrt 2 \\
\sqrt 2 & -1
\end{array}
\right).
\end{eqnarray}
We follow the representation basis in Refs.~\cite{Ishimori:2010au,Ishimori:2012zz}, which is different from one in Ref. \cite{Liu:2019khw}.

Explicitly, the modular form of weight 1 for the $T'$ doublet can be written by 
\begin{eqnarray}
\left(\begin{array}{c}
x_1 \\ y_1
\end{array}
\right)
= 
\left(
\begin{array}{c}
-\sqrt 2 \hat e_1 \\
 \hat e_1 +\frac13 \hat e_2
\end{array}
\right).
\end{eqnarray}
Their $q$-expansions can be written by 
\begin{eqnarray}
x_1 = -\sqrt 2 q^{1/3} + \cdots, \nonumber \\
y_1= 1/3 + \cdots.
\end{eqnarray}



\end{document}